\documentclass[floatfix,twocolumn,showpacs,aps,tightenlines,superscriptaddress]{revtex4-1}
\usepackage{graphicx}
\usepackage{color}
\usepackage{psfrag}
\usepackage{dcolumn}% Align table columns on decimal point
\usepackage{bm}% bold math

\usepackage{amssymb}
\usepackage{amsmath}
\usepackage{amsfonts}
\usepackage{longtable}
\usepackage{xspace}
\usepackage{verbatim}

\newcommand{\beq}{\begin{equation}}
\newcommand{\eeq}{\end{equation}}

\newcommand{\be}{\begin{equation}}
\newcommand{\ee}{\end{equation}}
\newcommand{\bea}{\begin{eqnarray}}
\newcommand{\eea}{\end{eqnarray}}
\newcommand{\bes}{\begin{subequations}}
\newcommand{\ees}{\end{subequations}}

\newcommand{\ud}{\mathrm{d}}
\newcommand{\MP}{{moving punctures}\xspace}
\newcommand{\MPA}{{moving punctures approach}\xspace}
\usepackage{bm}

\begin{document}

\title{High Energy Collisions of Black Holes Numerically Revisited}

\author{James Healy}
\affiliation{Center for Computational Relativity and Gravitation,
School of Mathematical Sciences,
Rochester Institute of Technology, 85 Lomb Memorial Drive, Rochester,
 New York 14623}
 
\author{Ian Ruchlin}
\affiliation{Center for Computational Relativity and Gravitation,
School of Mathematical Sciences,
Rochester Institute of Technology, 85 Lomb Memorial Drive, Rochester,
 New York 14623}
\affiliation{Department of Mathematics,
West Virginia University, Morgantown,
West Virginia 26506}

\author{Carlos O. Lousto}
\affiliation{Center for Computational Relativity and Gravitation,
School of Mathematical Sciences,
Rochester Institute of Technology, 85 Lomb Memorial Drive, Rochester,
 New York 14623}

\author{Yosef Zlochower}
\affiliation{Center for Computational Relativity and Gravitation,
School of Mathematical Sciences,
Rochester Institute of Technology, 85 Lomb Memorial Drive, Rochester,
 New York 14623}

\date{\today}

\begin{abstract}
We use fully nonlinear numerical relativity techniques to study high 
energy head-on collision of nonspinning, equal-mass
black holes to estimate the maximum gravitational radiation emitted
by these systems. Our simulations include improvements in the
construction of initial data, subsequent full 
numerical evolutions, and the computation of waveforms at infinity.
The new initial data
significantly reduces the spurious radiation
content, allowing for initial speeds much closer to  the speed of light,
i.e. $v\sim0.99c$. Using these new techniques, We  estimate the maximum radiated
energy from head-on collisions to be $E_{\text{max}}/M_{\text{ADM}}=0.13\pm0.01$.
This value differs from the second-order perturbative $(0.164)$ and 
zero-frequency-limit $(0.17)$ analytic computations, but is close
to those obtained by thermodynamic arguments $(0.134)$ and by 
previous numerical estimates $(0.14\pm0.03)$.
\end{abstract}

 \maketitle

\section{Introduction}\label{sec:intro}

The study of the high energy collision of two black holes is of
interest from both the theoretical point of view, to understand gravity
in its most extreme regime, and experimentally, since increasingly
high energy particle collisions could eventually have a non-negligible
probability for generating black hole pairs (see Ref.\@~\cite{Cardoso:2012qm} for a review).

The production of gravitational waves and the properties
of the final remnant after the collision of two black holes
has been the subject of theoretical study for over
half a century, with notable results such as
the area theorems by Hawking and
Penrose~\cite{Hawking71a,Eardley:2002re} and their application to bounds
on the energy radiated via gravitational waves. 
For instance, they find an upper bound for the  maximum energy radiated 
from a head-on collision of nonspinning black holes of $29\%$ of the
total mass.

More detailed estimates of the radiated energy have been computed
by applying perturbation theory~\cite{D'Eath:1992qu}
to the collision of ultrarelativistic black holes represented 
by shock waves~\cite{Aichelburg:1970dh}. 
Those computations reduce the above bound
to $25\%$ (when only including first-order corrections) and to
$16.4\%$ (when second-order corrections are included.
A $D-$dimensional generalization of the first-order computation~\cite{Coelho:2014gma} found that the proportion of energy radiated to
the initial mass scales as $1/2-1/D$.

Fully nonlinear numerical simulations of such collisions are now possible
thanks to the breakthroughs in numerical relativity~\cite{Pretorius:2005gq,Campanelli:2005dd,Baker:2005vv}.
The first full numerical study of the head-on collision of black
holes~\cite{Sperhake:2008ga} found a maximum 
efficiency of $14\pm3\%$. Those studies have been extended to 
grazing collisions~\cite{Sperhake:2009jz}, leading to an estimate of 
$35\%$  for the maximum energy radiated at a critical impact parameter.
Further studies including boson stars~\cite{Choptuik:2009ww}, fluid stars
\cite{East:2012mb,Rezzolla:2012nr},
black hole spins~\cite{Sperhake:2012me} and
unequal mass binaries~\cite{Sperhake:2015siy}
show that, at high energies, the structure 
(i.e. matter, spins and mass ratios) of the 
holes tends to be irrelevant for the collision outcomes.

The latest analytical computations of the energy radiated by
the head-on collision of two, equal mass, nonspinning black holes include
an estimate of $13.4\%$ based on black hole thermodynamics
arguments~\cite{Siino:2009vw} and $17\%$ based on a multipolar analysis of
the zero-frequency-limit (ZFL) approach~\cite{Berti:2010ce}.

In this paper we revisit the full numerical head-on computation
incorporating new techniques that improve the accuracy
of the simulations. These techniques include new initial data
with reduced spurious radiation content~\cite{Ruchlin:2014zva}, 
improved extraction
techniques with second order perturbative extrapolation~\cite{Nakano:2015pta},
and the use of new gauges~\cite{Alcubierre02b} and 
evolution systems~\cite{Alic:2011gg} in the
moving puncture approach~\cite{Campanelli:2005dd}.

We use the following standard conventions throughout this paper.
In all cases we use geometric units where $G=1$ and $c=1$. 
Latin letters ($i$, $j$, $\cdots$) represent spatial indices.
Spatial 3-metrics are denoted by $\gamma_{ij}$ and extrinsic
curvatures by $K_{ij}$. The trace-free part of the extrinsic curvature
is denoted by $A_{ij}$. A tilde indicates a conformally related
quantity. Thus $\gamma_{ij} = \psi^4 \tilde \gamma_{ij}$ and $A_{ij} =
\psi^{-2} \tilde A_{ij}$, where $\psi$ is some conformal factor. We
denote the covariant derivative associated with $\gamma_{ij}$ by $D_i$
and the covariant derivative associated with $\tilde \gamma_{ij}$ by
$\tilde D_i$. A lapse function is denoted by $\alpha$, while a shift
vector by $\beta^i$.

\section{Numerical Techniques}\label{sec:techniques}

\subsection{Initial Data}\label{sec:ID}

We use an extended version \textsc{TwoPunctures}~\cite{Ansorg:2004ds} thorn to generate 
puncture initial data~\cite{Ruchlin:2014zva} for boosted
black hole binary simulations.
In the conformal transverse-traceless (CTT) 
formalism~\cite{York99,Cook:2000vr,Pfeiffer:2002iy,AlcubierreBook2008}, 
the constraints on the initial spatial hypersurface $\Sigma_0$
become a set of elliptic differential equations 
for the conformal factor and potential vector (see Eq.~\eqref{eq:constraints}
below) through a conformal transformation
\begin{equation*}
  \gamma_{i j} = \psi^4 \tilde{\gamma}_{i j} \; .
\end{equation*}
We call $\tilde{\gamma}_{i j}$ the conformally related metric tensor. All objects with a tilde are associated with $\tilde{\gamma}_{i j}$.

As in Ref.~\cite{Ruchlin:2014zva}, to calculate the spatial metric and
extrinsic curvature associated with a boosted
black hole of mass $m$ and arbitrary linear $3$-momentum $P^i$, we
Lorentz boost the 4-dimensional Schwarzschild line element in isotropic 
Cartesian coordinates.
We then extract from the transformed metric the spatial metric
$\gamma_{i j}^*$, the lapse function $\alpha^*$, and the shift vector
$\beta^{i}_*$ (a super/subscript $*$ indicates that this is a single
black hole quantity). We 
then obtain the extrinsic curvature $K_{i j}^*$ on $\Sigma_{0}$ using the evolution equation for the spatial metric
\begin{equation*}
  K_{i j}^* = \frac{1}{2 \alpha^*} \left (D_{i}^* \beta_{j}^* +
  D_{j}^* \beta_{i}^* - \partial_{t'} \gamma_{i j}^* \right ) \; .
\end{equation*}
CTT separates this into trace and trace-free parts
\begin{equation*}
  K_{i j}^* = \psi^{-2}_* \tilde{A}_{i j}^* + \frac{1}{3} \psi^{4}_*
  \tilde{\gamma}_{i j}^* K^* \; ,
\end{equation*}
where $K^* = \gamma^{i j}_* K_{i j}^*$. For the conformal factor, we
make the standard choice
\begin{equation}\label{eq:psi}
  \psi^* = 1 + \frac{m}{2 r} \; ,
\end{equation}
where $r$ is the unboosted isotropic radius.

For example, if the boosted coordinates are given by
\begin{align}
  t' &= \gamma t + \gamma v y \; ,\\
  x' &= x \; , \\
  y' &= \gamma y + \gamma v t \; , \\
  z' &= z \; ,
\end{align}
then the conformal spatial line element on $\Sigma_{0}$ (defined by
$t'={\rm const}$) is given by
\begin{equation}
  \ud \tilde{\ell}^2 = \ud x'^2 + \gamma^2 \left [1 - \frac{16 (m - 2
  r)^2 r^4 v^2}{(m + 2 r)^6} \right ] \, \ud y'^2 + \ud z'^2 \; ,
\end{equation}
where $v$ is the magnitude of the local velocity vector
\begin{equation}
  v^i = \frac{P^i}{\sqrt{m^2 + P^j P_j}}
\end{equation}
(here the boost is along the $y$-axis),
$\gamma = (1 - v^2)^{-1/2}$,
and $r=\sqrt{x^2+y^2+z^2} = \sqrt{x'^2 + \gamma^2 (y' - v t')^2 + z'^2}$.

Our black hole binary initial data is
constructed using a superposition of metric and extrinsic curvature 
terms derived from the above expressions. To distinguish contributions
for the two black holes, we replace the $*$ super/subscript above with a $+$
or $-$.

The trace-free part of the extrinsic curvature is split
into background terms $\tilde{M}_{i j}$ and a longitudinal
correction
term obtained from a vector $b_i$.
Here
\begin{equation}
  \tilde M_{ij} = \tilde A^{(+)}_{ij} + \tilde A^{(-)}_{ij},
\end{equation}
where $\tilde A^{(+)}_{ij}$ and $\tilde A^{(-)}_{ij}$ are the
trace-free part of the conformal
extrinsic curvature of a single boosted black holes located at $\vec r =
\vec r_+$ and $\vec r = \vec r_-$.
Note that the trace-free part of the
single boosted black hole extrinsic curvature
will have a small trace with respect to a metric constructed by 
superimposing  two different 
background metrics. We remove this extra trace term prior to
solving the initial data equations, i.e.,
$\tilde M_{ij} \to \tilde M_{ij} - \frac{1}{3}\tilde \gamma_{ij}
\tilde \gamma^{lm} \tilde
M_{lm}$ (where $\tilde \gamma_{ij}$ is the superimposed background
metric).
The complete trace-free part of the extrinsic curvature for the
superimposed spacetime is given
by 
\begin{equation}
  \tilde{A}_{i j} = \tilde{M}_{i j} + \frac{1}{\tilde{\alpha}} (\tilde{\mathbb{L}} b)_{i j} \; ,
\end{equation}
where $\alpha = \psi^{6} \tilde{\alpha}$ and $(\tilde{\mathbb{L}}
b)_{i j} \equiv \tilde{D}_{i} b_{j} + \tilde{D}_{j} b_{i} -
\frac{2}{3} \tilde{\gamma}_{i j} \tilde{D}_{k} b^{k}$ is the longitudinal vector gradient. As part of the freely specifiable parameters we set $\tilde{\alpha} = 1$.

In the puncture approach, we write the conformal factor as singular parts plus a finite correction, $u$,
\begin{equation}
  \psi = \psi_{(+)} + \psi_{(-)} - 1 + u \; ,
\end{equation}
where $\psi_{(\pm)}$ are the conformal factors~\eqref{eq:psi}
associated with the individual, isolated black holes located at 
positions labeled as $(+)$ and $(-)$,
with spatial metric tensors $\tilde{\gamma}_{i j}^{(\pm)}$.

Given these choices, the Hamiltonian and momentum constraints become equations for the correction functions $u$ and $b^i$
\begin{subequations}
  \label{eq:constraints}
  \begin{align}
    \tilde{D}^2 u - \frac{\psi \tilde{R}}{8} - \frac{\psi^{5} K^2}{12}
    + \frac{\tilde{A}_{i j} \tilde{A}^{i j}}{8 \psi^{7}} + \tilde{D}^2\left(\psi_{(+)} + \psi_{(-)}\right) &= 0 \; , \\
    \tilde{\Delta}_{\mathbb{L}} b^i + \tilde{D}_{j} \tilde{M}^{i
    j} - \frac{2}{3} \psi^{6} \tilde{\gamma}^{i j} \tilde{D}_{j}
    K &= 0 \; ,\label{eq:momentum}
  \end{align}
\end{subequations}
where $\tilde{\Delta}_{\mathbb{L}} b^i \equiv \tilde{D}_{j}
(\tilde{\mathbb{L}} b)^{i j}$ is the vector Laplacian and $\tilde{R}$
is the scalar curvature associated with $\tilde{\gamma}_{i j}$. The
solutions are required to obey Dirichlet conditions at infinity
\begin{equation*}
  \lim_{r \to \infty} u = 0 \quad \text{and} \quad \lim_{r \to \infty} b^i = 0 \; .
\end{equation*}

In order to deal with the puncture singularities, 
we introduce attenuation functions to both modify the background
metric and mean curvature, as well as modify the singular source terms
inside the horizons themselves.
The first type of attenuation, which is consistent with the
constraints everywhere, is used in the superposition of the background
conformal metrics and has the form,
\begin{equation*}
  \tilde{\gamma}_{i j} = \delta_{i j} + f_{(+)} \left(\tilde{\gamma}_{i j}^{(+)} - \delta_{i j}\right) + f_{(-)} \left(\tilde{\gamma}_{i j}^{(-)} - \delta_{i j}\right) \; ,
\end{equation*}
where
\begin{equation*}
  f_{(\pm)} = 1 - e^{-(r_{(\mp)}/\omega_{(\pm)})^p} \; ,
\end{equation*}
and $r_{(\pm)}$ is the coordinate distance from 
a field point to the location of puncture $(\pm)$. 
The parameters $\omega_{(\pm)}$ control the 
steepness of the attenuation.  We take the smallest possible 
power index $p = 4$ to achieve convergence of the solutions to 
the constraints.

The second attenuation function is used to modify the background mean
curvature  and the
source term in the momentum constraint equations.
This takes the form
\begin{align*}
  K &= f_{(+)} g K_{(+)} + f_{(-)} g K_{(-)} \; ,\\
  \tilde{D}_{i} \tilde{M}^{ij} &= g  \tilde{D}_{i} \tilde{A}^{i j}_{(+)} + 
  g \tilde{D}_{i} \tilde{A}^{i j}_{(-)},
\end{align*}
where 
\begin{align*}
g &= g_{+}\times g_{-} \; ,\\
  g_{\pm} &= 
          \begin{cases} 
     1 & \mbox{if } r_\pm > r_{\rm max} \\
     0 & \mbox{if } r_\pm < r_{\rm min} \\
            {\cal G}(r_{\pm}) & \mbox{otherwise},
  \end{cases} \; ,\\
  {\cal G}(r_\pm) &= \frac{1}{2}\left[1+ \tanh\left(\tan\left[ \frac{\pi}{2}
  \left(-1 + 2 \frac{r_{\pm}-r_{\rm min}}{r_{\rm max} -
  r_{\rm min}}\right)\right]\right)\right],
\end{align*}
and the parameters $r_{\rm min} < r_{\rm max}$ are chosen
to be within the horizon.
Note that the $g$ attenuation function is used to modify the
constraint equations themselves inside the horizon. We will refer to
the above data as the {\it standard data} in the sections below.

In addition, we consider a second type of initial data closely related
to the above approach. For this, which we shall
refer to as {\it approximate data} in the sections below, 
we analytically remove the singularity associated
with
$\tilde D_{i} \tilde M^{ij}$ by making the following two
approximations.
First, we take $\tilde M_{ij}$ to be the sum of the
two Kerr trace-free extrinsic curvatures without correcting for the
fact that the background metric is now a superimposed metric. Second,
in the source term of Eq.~(\ref{eq:momentum}), we replace 
$\tilde{D}_{i} \tilde{A}_{\pm}^{ij}$ with $(\tilde{D}_{i} - \tilde{D}^{\pm}_i)\tilde{A}_{\pm}^{ij} +
\tilde{D}^{\pm}_{i} \tilde{A}_{\pm}^{ij}$, where $\tilde{D}^{\pm}_{i}$ and
$\tilde A_{\pm}^{ij}$ are the covariant derivative and extrinsic
curvature associated with the two background conformal Kerr metrics. The former term contains no derivatives of
$\tilde A_{\pm}^{ij}$, while the latter is evaluated analytically.
This is an additional approximation because we neglect the
fact that indices in $\tilde A_{\pm}^{ij}$ are actually raised using the full
superposed metric, rather than the associated Kerr metric.
The net result of these approximations
is that the initial data solution converges (relatively) rapidly with
collocation points, but the resulting constraints converge
to a {\it} small non-zero value. We use a subsequent CCZ4 evolution to
remove this residual violation. This allow us to quantify the effects of
small violations of the initial constraints and how to control them.

\subsection{Convergence of Initial Data}\label{sec:ID_conv}
\begin{figure*}
  \includegraphics[height=.4\textwidth, angle=270]{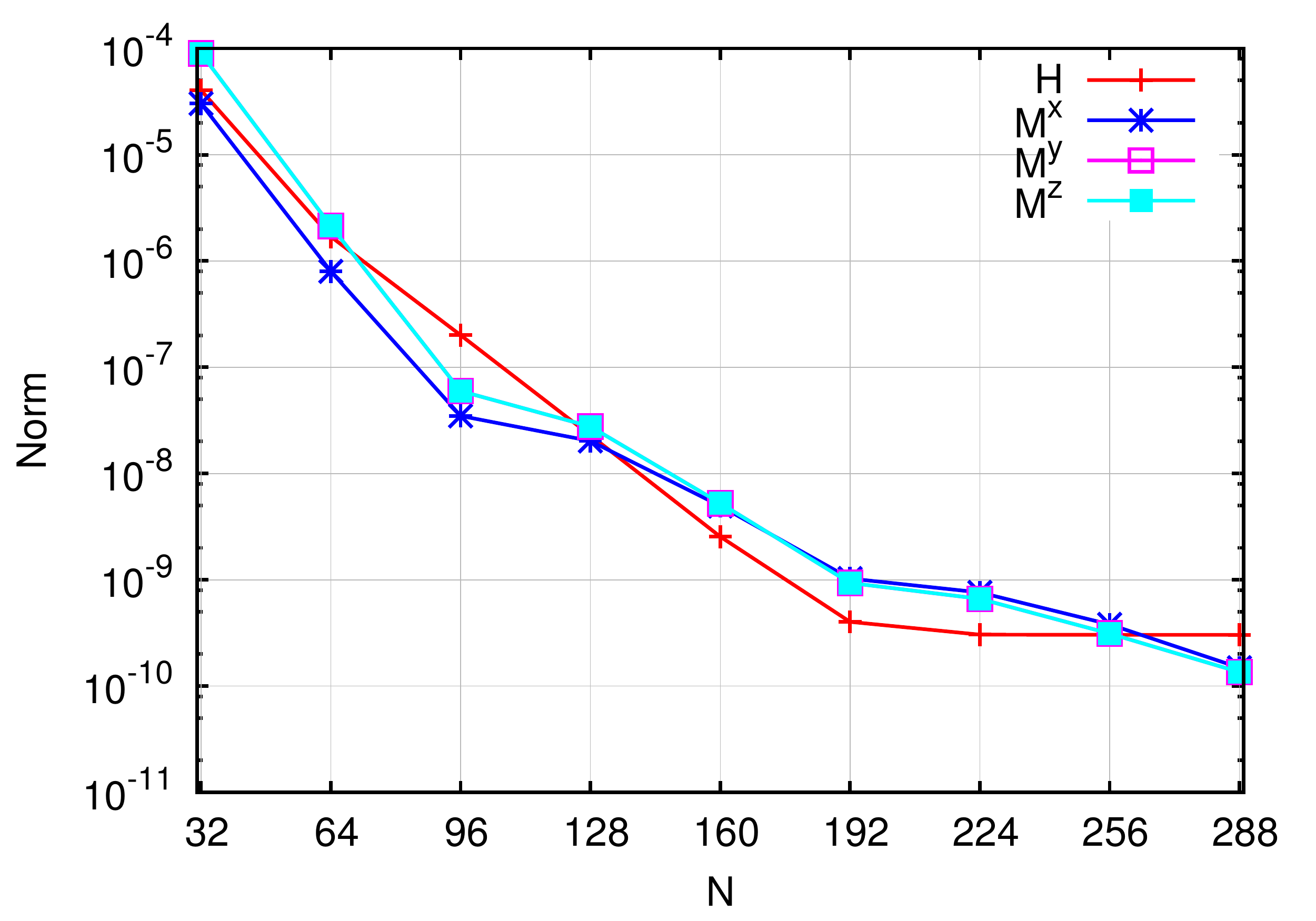}
  \includegraphics[height=.4\textwidth, angle=270]{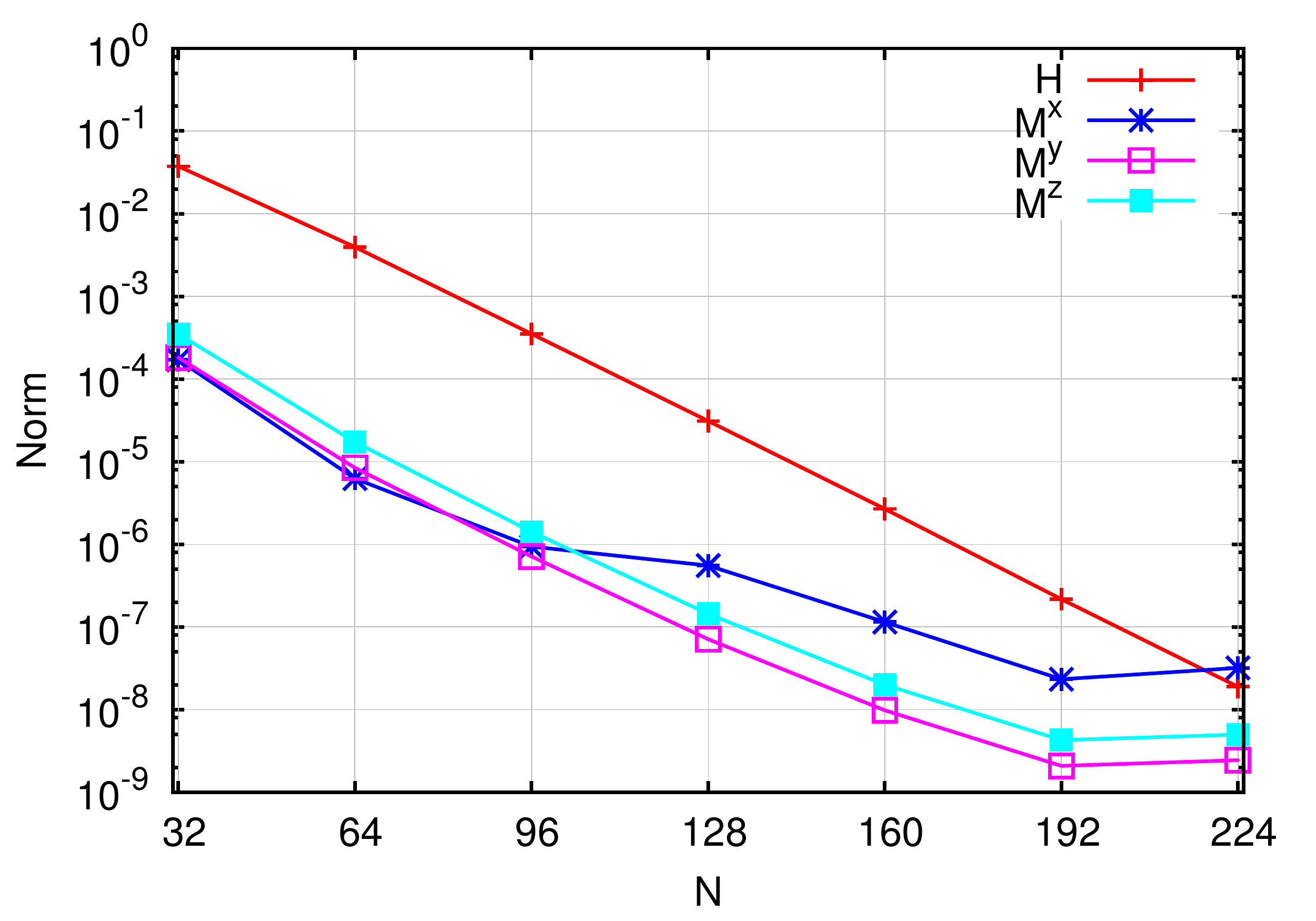}\\

  \includegraphics[height=.4\textwidth, angle=270]{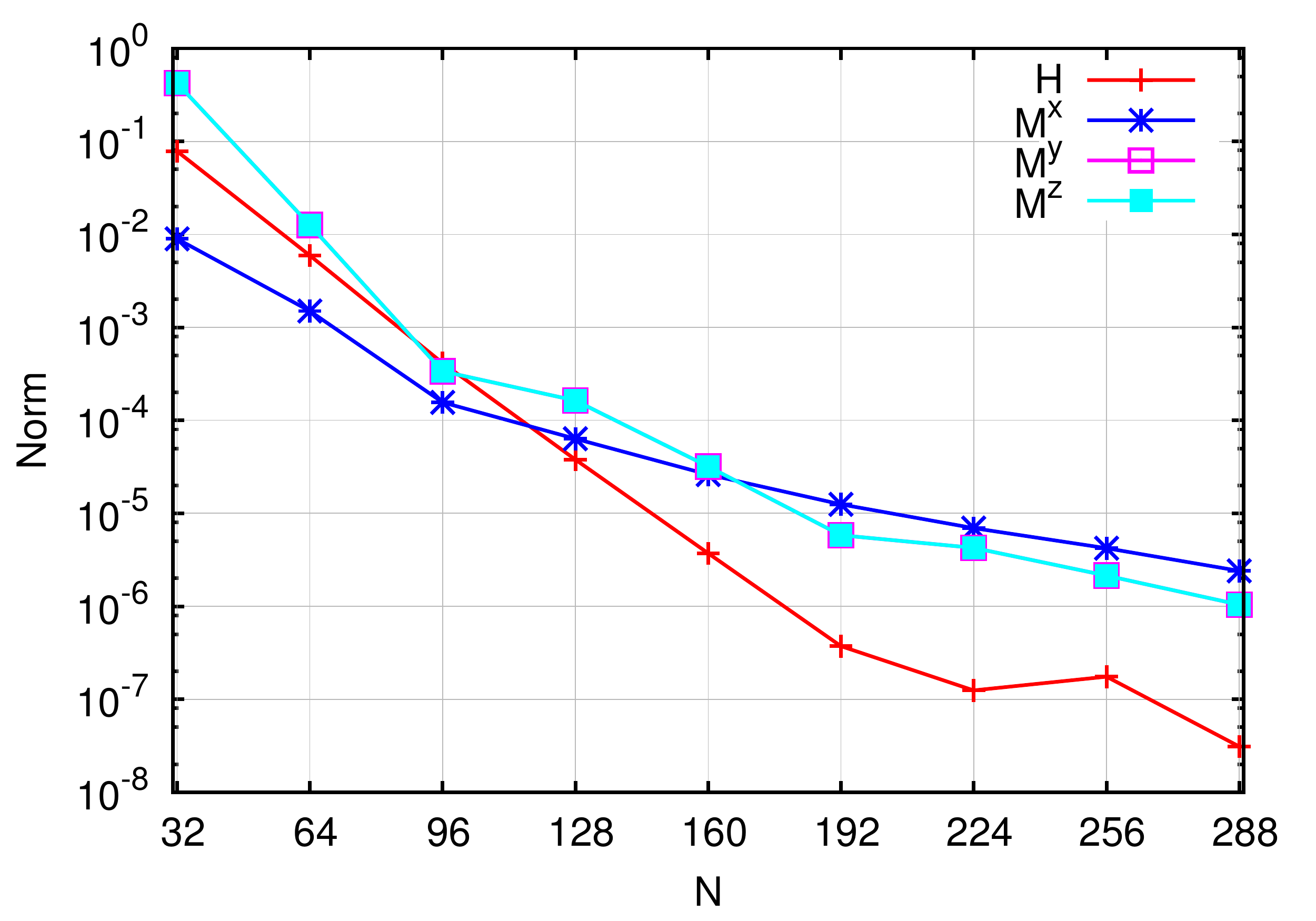}
  \includegraphics[height=.4\textwidth, angle=270]{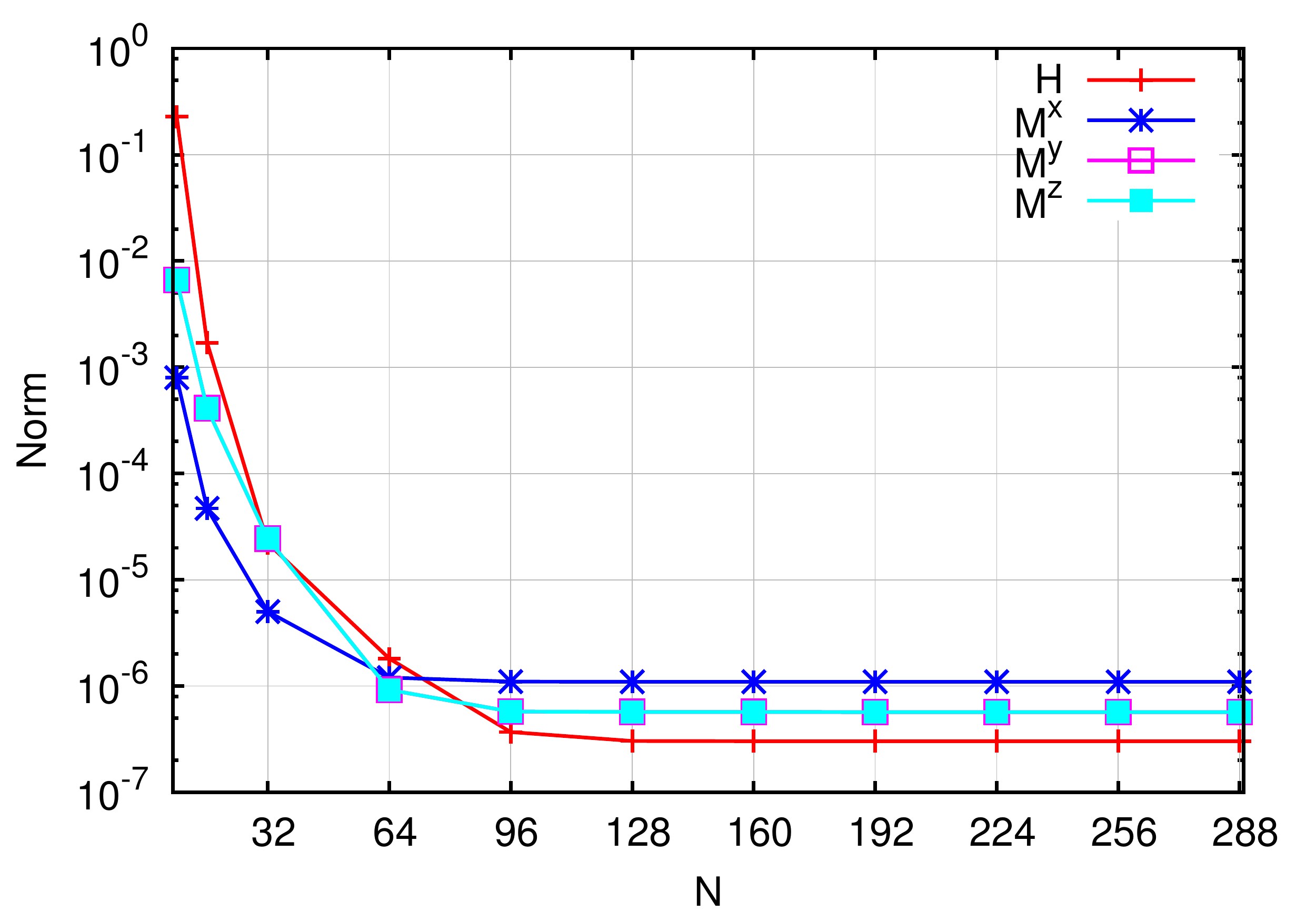}
  \caption{The $L^2$-norms of the Hamiltonian and momentum constraints for
    the {\it standard data} over
    the full grid extending to $400M$ (top-left), over a small volume 
    centered on $(x,y,z)=(5.0, 0.35, 0.75)$ (top-right),  over a small volume 
    containing the $x$-axis centered on $(4.5, 0, 0)$ (bottom-left).
    The number of collocation points is given by
$N\times N\times 32$ in the $A$, $B$, and $\phi$ dimensions,
respectively. The bottom-right plot shows the constraint violation over  the
full grid with the {\it approximate data} instead.
\label{fig:id_conv}}
\end{figure*}

\begin{figure*}
  \includegraphics[width=.45\textwidth,angle=0]{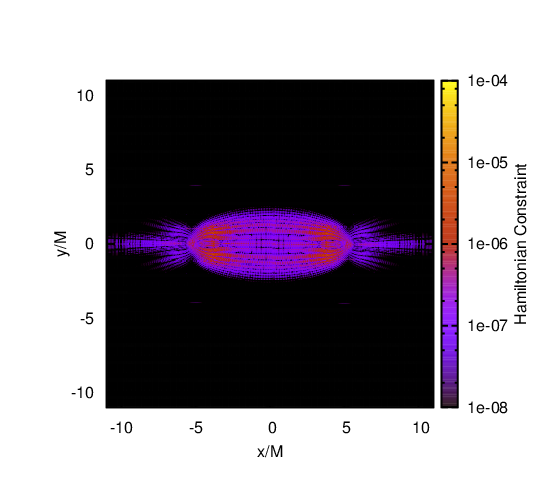}
  \includegraphics[width=.45\textwidth,angle=0]{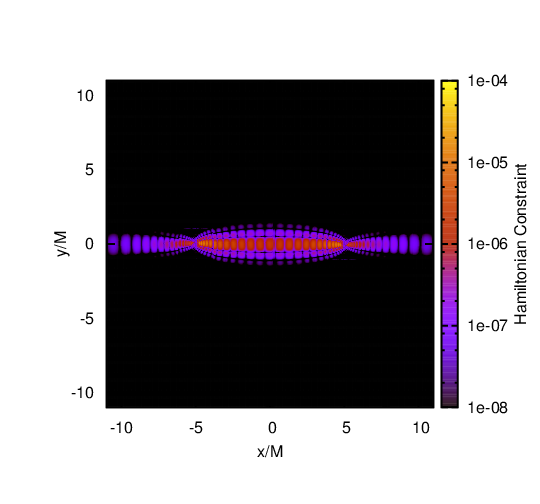}

  \includegraphics[height=.4\textwidth,angle=270]{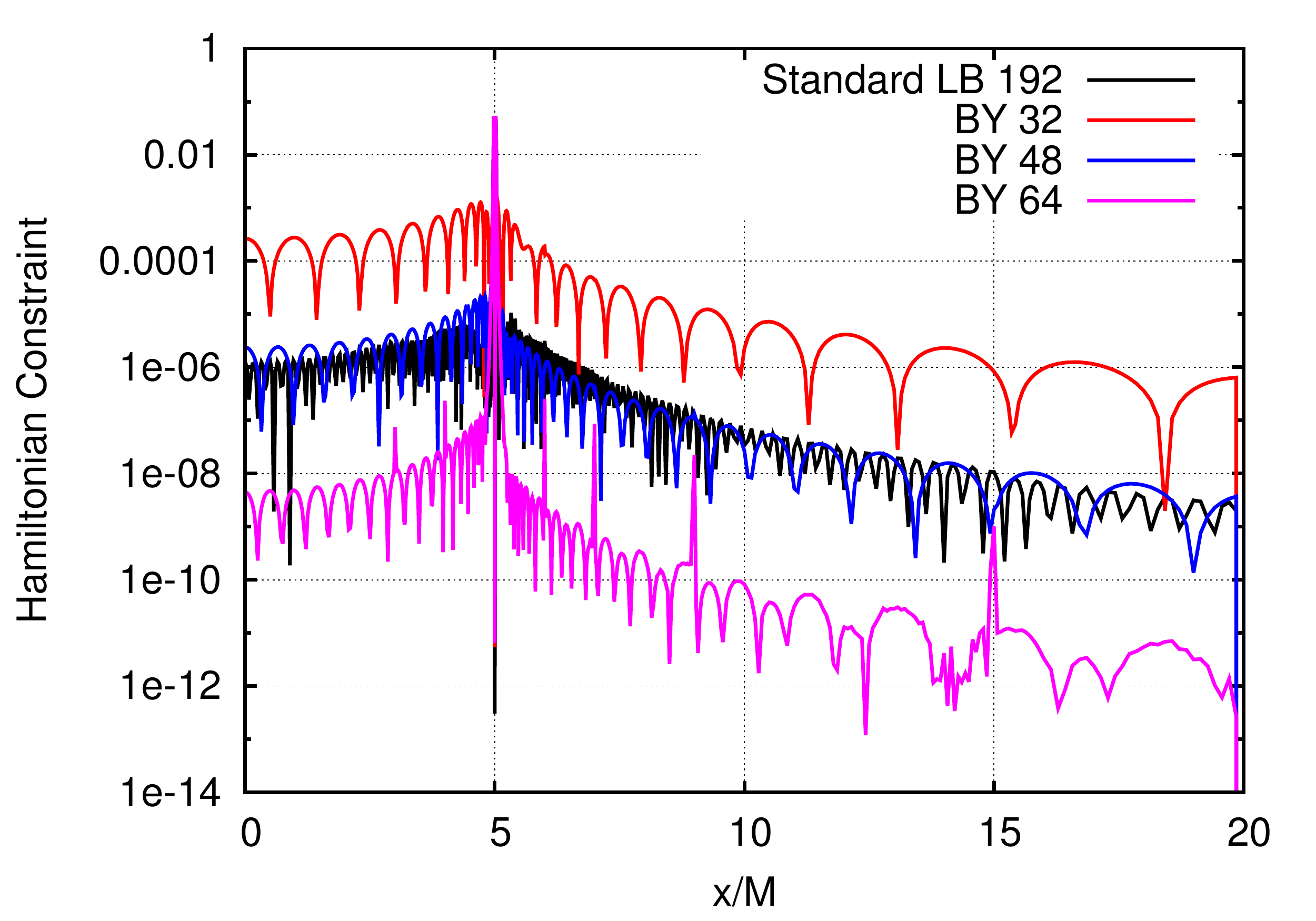}
  \includegraphics[height=.4\textwidth,angle=270]{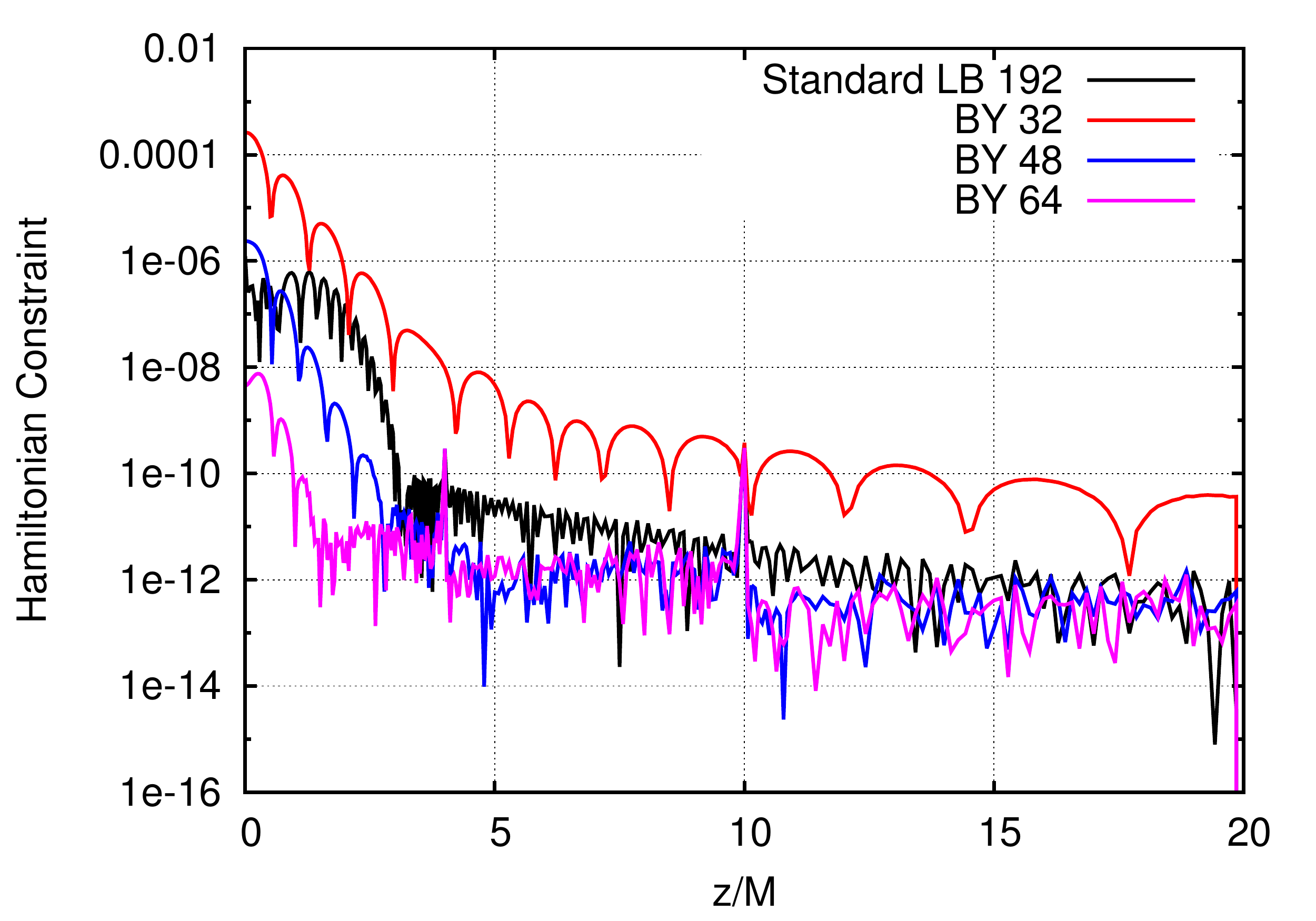}
  \caption{The Hamiltonian constraint in the $xy$-plane of the {\it standard data} 
    for $P/m_{\text{irr}}=1$ with initial separation $d/M=10$ 
    in the region around the black holes for $N=192$ collocation points (Top left). For comparison we also show the violations of the Hamiltinian constraint for BY data for $N=48$ collocation points (Top right).
In the lower panels we show the violation of the Hamiltonian constraint along
the axis containing the black holes and perpendicular to it for our data and for BY data for different number of colocation points $N=32,48,64$.}
    \label{fig:id_2d_HC}
\end{figure*}

As shown in Fig.~\ref{fig:id_conv}, we verified the exponential
convergence of the constraint violations of the initial data (with
collocation points) using the $L^2$-norms (RMS). We find that
volume-averaged constraint violation (i.e., $L^2$ over the entire
simulations domain) converge to levels of $\sim 5\times 10^{-10}$. The
constraint violations are largest near the $x$-axis (see
Fig.~\ref{fig:id_2d_HC}). To verify the convergence of the data there, we
calculated the $L^2$-norm over as small box of width $0.5M$ centered on
the $x$-axis. This box is chosen so that it lies just outside one of
the horizons. The momentum constraints converge exponentially, but at a
relatively slow rate, in this volume. The Hamiltonian converges
exponentially to a level of $\sim 10^{-7}$. The source of the
relatively large violations is a high-frequency component in the
initial data induced by the scale of the attenuation function.
Note that the convergence of the {\it approximate data} is much faster
with collocation points, but also converges to a non-zero value.

Figure~\ref{fig:id_2d_HC} shows the initial data Hamiltonian constraint violation
on the $xy$-plane for a configuration with $P/m_{\text{irr}}=1$ and initial separation $d/M=10$ . Note the high-frequency residual. By increasing the
width of the attenuation function $g$ above, we were able to partially mitigate
the high-frequency noise in the constraint residuals using 
$N=192^2\times4$ collocation points. For comparison we also
show the same configuration with  $P/m_{\text{irr}}=1$ and initial separation $d/M=10$ but for $N=48^2\times4$ collocation points, which represents a medium
resolution for BY data. The choice of a lower number of collocation points
($N=48^2\times4$) for BY is because the BY system is algebraically
simpler than the new data (among others, the momentum 
constraints are solved exactly, and the background is flat). We thus expect that for a given number 
of collocation points, BY data will have a much smaller constraint
violation, which is indeed what we see (see bottom panels of this
figure).  From the figure, we see  that we
can reach acceptable levels of constraint violations with our data, but
require a much larger number of collocation points than for BY.

\subsection{Evolution}\label{sec:evolution}

We evolve black hole binary initial data sets using the 
{\sc LazEv}~\cite{Zlochower:2005bj} implementation of the \MPA 
for both the BSSNOK formalism~\cite{Nakamura87, Shibata95,
Baumgarte99} and the conformal and covariant formulation of the Z4 (CCZ4) system
(Ref.~\cite{Alic:2011gg}) which includes stronger damping of
the constraint violations than the BSSNOK system.
For the runs presented here, we use
centered, eighth-order accurate finite differencing in
space~\cite{Lousto:2007rj} and a fourth-order Runge-Kutta time
integrator. Our code
uses the {\sc Cactus}/{\sc EinsteinToolkit}~\cite{cactus_web,
einsteintoolkit} infrastructure.  We use the {\sc Carpet} mesh 
refinement driver to provide a ``moving boxes'' style of mesh refinement
\cite{Schnetter-etal-03b}.  Fifth order Kreiss-Oliger dissipation is added to
evolved variables with dissipation coefficient $\epsilon=0.1$.
Note that when using CCZ4, we chose damping parameters
$\kappa_1 = 0.1$, $\kappa_2=0$, and $\kappa_3=0$
(see~\cite{Alic:2011gg}).

We locate the apparent horizons using the {\sc AHFinderDirect}
code~\cite{Thornburg2003:AH-finding} and measure the horizon spins
using the isolated horizon (IH) algorithm~\cite{Dreyer02a}.
To compute the radiated angular momentum components, we
use formulas based on ``flux-linkages''~\cite{Winicour_AMGR}, explicitly written in terms of $\Psi_4$~\cite{Campanelli:1998jv, Lousto:2007mh}. We then extrapolate
those extractions to an infinite observer location using
formulae accurate to $\mathcal{O}(1/r^2_{\text{obs}})$~\cite{Nakano:2015pta}.

We obtain accurate, convergent waveforms and horizon parameters by
evolving this system in conjunction with a modified 1+log lapse and a
modified Gamma-driver shift condition~\cite{Alcubierre02a,
Campanelli:2005dd, vanMeter:2006vi}. The lapse and shift are evolved with
\begin{subequations}
  \label{eq:gauge}
  \begin{align}
    (\partial_t - \beta^i \partial_i) \alpha &= - \alpha^2 f(\alpha) K \; , \\
    \partial_t \beta^a &= \frac{3}{4} \tilde{\Gamma}^a - \eta \beta^a \; .
  \end{align}
\end{subequations}
where $\eta=2$.

We have found that the choice $f(\alpha)=8/(3\alpha(3-\alpha))$ 
(approximate shock avoiding~\cite{Alcubierre02b}) proves to be more stable and
convenient when dealing with highly boosted moving punctures at relatively
short separations, $\approx~100M$ (this proved particularly useful for
the CCZ4 simulations described below).
This is due to the fact that the shock avoiding gauge suppresses a
large amplitude gauge wave that would otherwise be focused by the
black holes and subsequently trigger a Courant
violation when the lapse gets too big.
For the initial form of the lapse we use
$\alpha(t=0)=1/(2\psi_{\text{BL}}-1)$, where   
$\psi_{\text{BL}}=1+m_{(+)}/(2r_{(+)})+m_{(-)}/(2r_{(-)})$. This proved
to produce more accurate evolutions for highly spinning black
holes~\cite{Ruchlin:2014zva} and we will also adopt it for
the highly boosted cases in this paper.

For both sets of evolutions, CCZ4 and BSSNOK, there are between 11 
and 13 levels of mesh refinement depending on the momentum
of the black holes. Since we start the initial separations of the CCZ4 
simulations much 
farther apart than the BSSNOK evolutions, the coarsest levels of the
grid structure differ between the two sets of evolution.  The CCZ4
evolutions have an outer boundary of $800M$, while the BSSNOK 
evolutions have an outer boundary of $400M$.  The finest levels are 
the same for both sets of evolutions.  We label the different resolution runs
by $nX$ where $X$ is a global grid factor.  For all evolution runs,
we use a grid of $n120$.  For this case, the finest resolution 
for the $P/m_{\text{irr}}=0.3$ case is $M/307.2$ and finest resolution 
for the $P/m_{\text{irr}}=4.0$ case is $M/1228.8$.  Full details of the $n120$ grid
structure is given in Table \ref{tab:grid}. 

\begin{table}
  \caption{Table of grid structure  for case $n120$. For $P/m_{\text{irr}}$ up to 2 we 
   use up to mesh level 10.  For $P/m_{\text{irr}}=3$ we include an additional 
   level and for $P/m_{\text{irr}}=4$ we use all 13 mesh refinement levels. 
   Whether the refinement level's grid is centered on the origin or around the black holes (BHs)
   is given in column 2.  The radius of the box is given in column 3.  For meshes
   with two values, the first is for the BSSNOK evolution of the {\it
   standard data}, and the second is
   for the CCZ4 evolutions of the {\it approximate data}.
  }\label{tab:grid}
\begin{ruledtabular}
\begin{tabular}{l|ccc}
\hline
Mesh Number & Centered on & Radius & Resolution\\
\hline
0   & Origin &  400,800 & $M/0.3$\\
1   & Origin &  200,500 & $M/0.6$\\
2   & Origin &  140,300 & $M/1.2$\\
3   & BHs    &  32      & $M/2.4$\\
4   & BHs    &  16      & $M/4.8$\\
5   & BHs    &  8       & $M/9.6$\\
6   & BHs    &  4       & $M/19.2$\\
7   & BHs    &  2       & $M/38.4$\\
8   & BHs    &  1.2     & $M/76.8$\\
9   & BHs    &  0.6     & $M/153.6$\\
10  & BHs    &  0.3     & $M/307.2$\\
11  & BHs    &  0.15    & $M/614.4$\\
12  & BHs    &  0.08    & $M/1228.8$\\
\hline

\end{tabular}
\end{ruledtabular}
\end{table}

\begin{figure}
  \includegraphics[angle=270,width=\columnwidth]{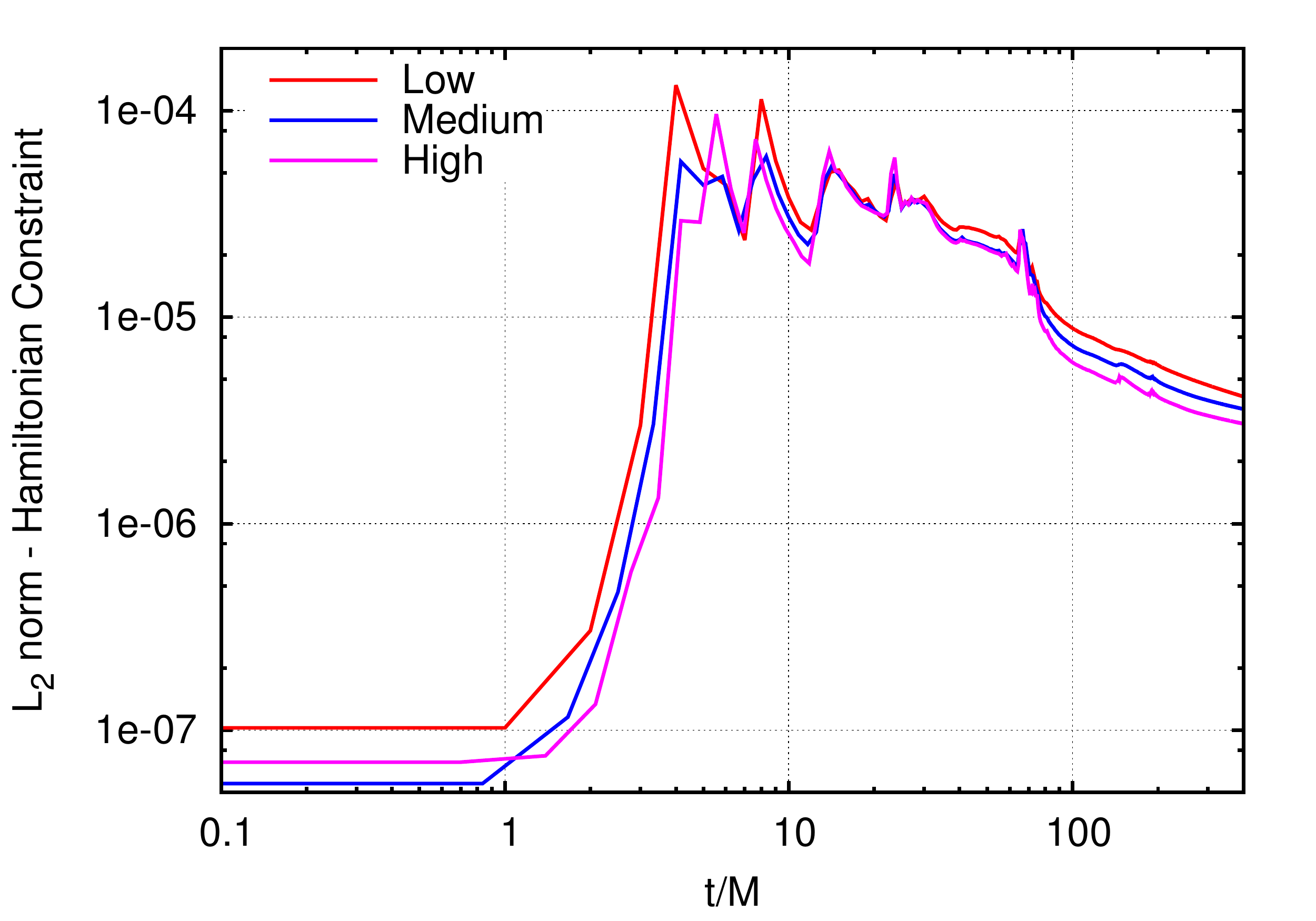}
  \caption{The $L^2$-norm of the Hamiltonian constraint violation
    versus time for a $P/m_{\rm irr}=2$ simulations 
using {\it standard data} evolved with BSSNOK at three different 
resolutions.  At early times, the constraint violations are much
smaller, but the violations become much larger than their initial
values for most of the simulation. 
Globally the constraints decrease with resolution only slightly due to 
high-frequency noise on the grid.}\label{fig:const_v_time}
\end{figure}

Figure~\ref{fig:const_v_time} shows the constraint violations versus
time for a $P/m_{\rm irr} =2$ simulation using BSSNOK evolutions of {\it
standard data} for three resolutions $(n100,\ n120,\ n144)$. During
most of the run, after the initial settling of gauges, 
the constraint violations are convergent. We observe a hyperconvergent
(4th-8th) order for the merger phase and then an slower convergence
(due to residual grid interboundary radiation) of nearly 
first order after merger.

\section{Relativistic head-on collisions}\label{sec:HOC}

\begin{figure}
\includegraphics[angle=270,width=\columnwidth]{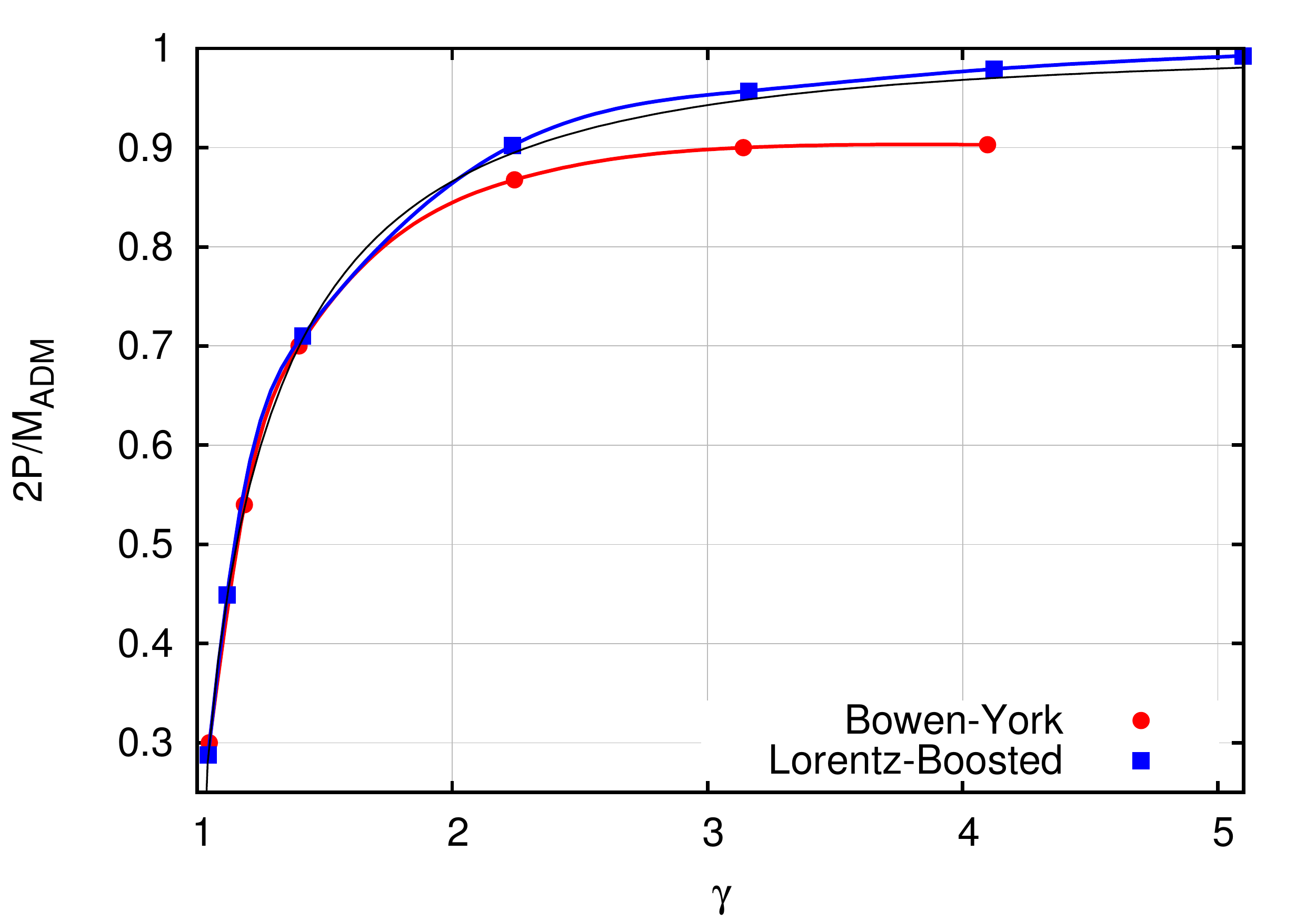}
\caption{The center of mass speeds of the black holes for
the Lorentz boosted data in this paper and Bowen-York data.
The latter displays a limitation, reaching speeds of only $v<0.9c$,
while the former can reach near the ultrarelativistic regime.
The thin line represent the special relativistic speed expression
$v=c\sqrt{1-1/\gamma^2}$.
}
\label{fig:BBH_ID_trunc}
\end{figure}

A major difference between our work here and previous studies, see
Refs.~\cite{Sperhake:2008ga, Sperhake:2009jz,
 Sperhake:2012me,
Sperhake:2015siy}, is that we use non-conformally flat initial data.
The Bowen-York initial data used previously are
limited to representing black holes moving at speeds $v<0.9c$, as
shown in
Fig.~\ref{fig:BBH_ID_trunc}.
The reason for this is that the assumption of conformal flatness introduces a Brill wave that gets
stronger as the momentum parameter is increases. Most of this wave is
absorbed by the black holes, leading them to increase in mass
proportional to the momentum parameter. The net effect is that the
ratio of momentum to mass of each black hole can never be larger than
$P/m_{\rm irr}\sim 2$ (i.e.,
$v/c \sim 0.9$). Note that here we use the irreducible mass of each
black hole in place of the particle rest mass. 

The situation is similar to that observed in highly spinning
black holes, where the conformally flat ansatz for the 3-metric
leads to a limitation \cite{Cook90a, Dain:2002ee, Lousto:2012es}
in the maximum intrinsic spin of the black
hole of around $S/m^2\approx0.93$.

On the other hand, Fig.~\ref{fig:BBH_ID_trunc} shows that the 
new data we use here is not limited by
this condition and can reach velocities closer to the speed of
light, i.e. $v\sim0.99c$. This is due to the much lower initial
radiation content of the data. We will exploit
this characteristic of the initial data to obtain a more accurate
estimate of the maximum gravitational radiation produced by head-on
collision of two equal mass, nonspinning, black holes.

In order to explore the dependence of the radiated energy on
the magnitude of initial momentum of the two black holes, and then extrapolate 
the results to the ultrarelativistic 
limit, we set up a series of simulations with $P/m_{\rm irr}$ ranging from $0.3$
to $4.03$ (see Tables~\ref{tab:EradTab} and \ref{tab:EradTab2}). We have chosen a relatively large
initial separation of the black holes in order to ensure that the
isolated horizon formalism can be used to accurately measure the mass
of the black holes and to ensure that the momentum parameter used in
the simulations corresponds closely to the momentum of the black holes
at infinite separation. 

\begin{figure}
  \includegraphics[angle=270,width=\columnwidth]{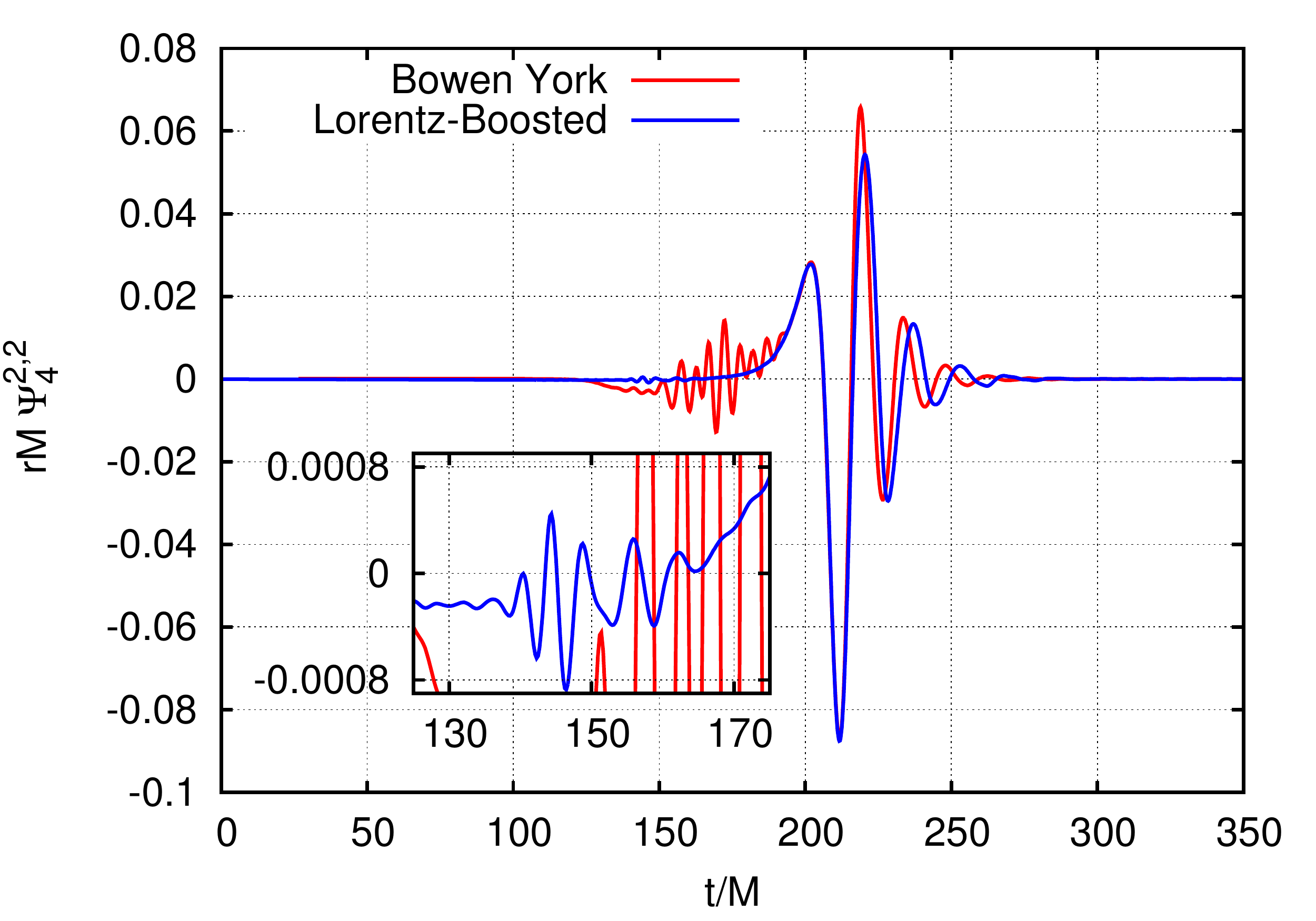}
  \includegraphics[angle=270,width=\columnwidth]{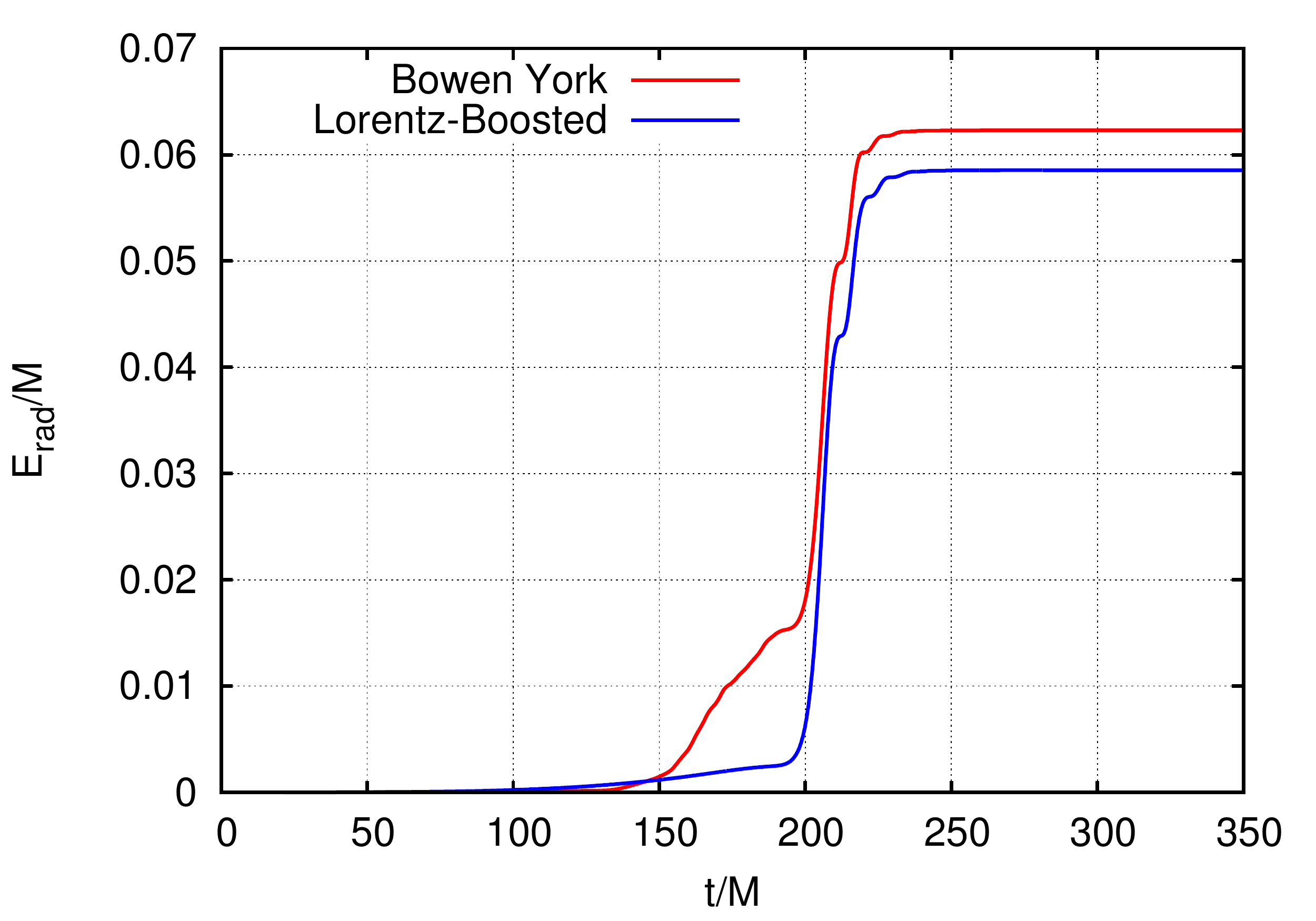}
  \caption{Top panel: The $(\ell=2,m=2)$ mode of $\psi_4$ for the $P/m_{\rm irr}=2$ Bowen-York
  and Lorentz-Boosted simulations. Note the spurious signal at $t\sim150M$.  The inset zooms in on 
  the spurious signal of the Lorentz-Boosted waveform.  Bottom Panel: The energy radiated for the two
  waveforms in the top panel. The contribution of the spurious
  radiation can be seen by looking at the energy radiated up to time
  $t\sim 200M$. 
  }\label{fig:psi4}
\end{figure}

In Fig.~\ref{fig:psi4}, we show the $(\ell=2, m=2)$ mode of $\psi_4$ for a
typical simulations (here $P/m_{\rm irr}=2$) as seen by an observer at
$r=130M$. The spurious radiation is evident
in the burst near $t\sim150M$, well before the merger signal. As can
be seen from the figure, for the new data, the spurious radiation
contains about $4\%$ of the total energy radiated for the {\it
standard data}. On the other hand, the Bowen-York spurious radiation
content is $24\%$ of the total.

\begin{table*}
\caption{Table of initial parameters and energy radiated for the
{\it standard initial data} evolved with BSSNOK. }\label{tab:EradTab}
\begin{ruledtabular}
\begin{tabular}{lcc|ccc|cc}
$P/M_{\text{ADM}}$ & $M_{\text{ADM}}/M$ & $m_{\text{irr}}/M_{\text{ADM}}$ & $P/m_{\text{irr}}$ & $\gamma$ & $d/M$ & $E_{\text{rad}}/M_{\text{ADM}}$ & $\delta E_{\text{rad}}/M_{\text{ADM}}$\\
\hline
0.1437 & 1.0008 & 0.4804 & 0.30 & 1.0438 & 100 & 0.0011 & 4.8e-5\\
0.2238 & 1.0028 & 0.4488 & 0.50 & 1.1174 & 100 & 0.0031 & 1.3e-6\\
0.3547 & 1.0093 & 0.3555 & 1.00 & 1.4126 & 100 & 0.0182 & 2.6e-4\\
0.4510 & 1.0177 & 0.2583 & 2.00 & 2.2336 & 100 & 0.0585 & 1.3e-3\\
0.4792 & 1.0268 & 0.1597 & 3.00 & 3.1630 & 100 & 0.0858 & 1.8e-3\\
0.4886 & 1.0250 & 0.1220 & 4.00 & 4.1272 & 150 & 0.0957 & 1.3e-4\\

\end{tabular}
\end{ruledtabular}
  \caption{Table of initial parameters and energy radiated for the
  {\it approximate data} evolved with CCZ4.  For each system, the
initial ADM mass is normalized to 1. }\label{tab:EradTab2}
\begin{ruledtabular}
\begin{tabular}{lcc|ccc|cc}
$P/M_{\text{ADM}}$ & $M_{\text{ADM}}/M$ & $m_{\text{irr}}/M_{\text{ADM}}$ & $P/m_{\text{irr}}$ & $\gamma$ & $d/M_{\text{ADM}}$ & $E_{\text{rad}}/M_{\text{ADM}}$ & $\delta E_{\text{rad}}/M_{\text{ADM}}$\\
\hline
0.1439 & 1.0000 & 0.4807 & 0.30 & 1.0440 & 100 & 0.0011 & 6.8e-7\\
0.2245 & 1.0000 &  0.4498 & 0.50 & 1.1180 & 100 & 0.0030 & 6.1e-6\\
0.3558 & 1.0000 & 0.3559 & 1.00 & 1.4142 & 200 & 0.0183 & 1.2e-4\\
0.4530 & 1.0000 & 0.2263 & 2.00 & 2.2361 & 200 & 0.0592 & 4.7e-4\\
0.4800 & 1.0000 & 0.1594 & 3.01 & 3.1717 & 300 & 0.0859 & 1.1e-3\\
0.4908 & 1.0000 & 0.1217 & 4.03 & 4.1231 & 400 & 0.0988 & 9.7e-4\\

\end{tabular}
\end{ruledtabular}

\end{table*}

To extrapolate the energy radiated to infinite observer location, we use
7 finite observers and extrapolate using a 1st order
and 2nd order polynomial. 

\begin{figure}
\includegraphics[angle=270,width=\columnwidth]{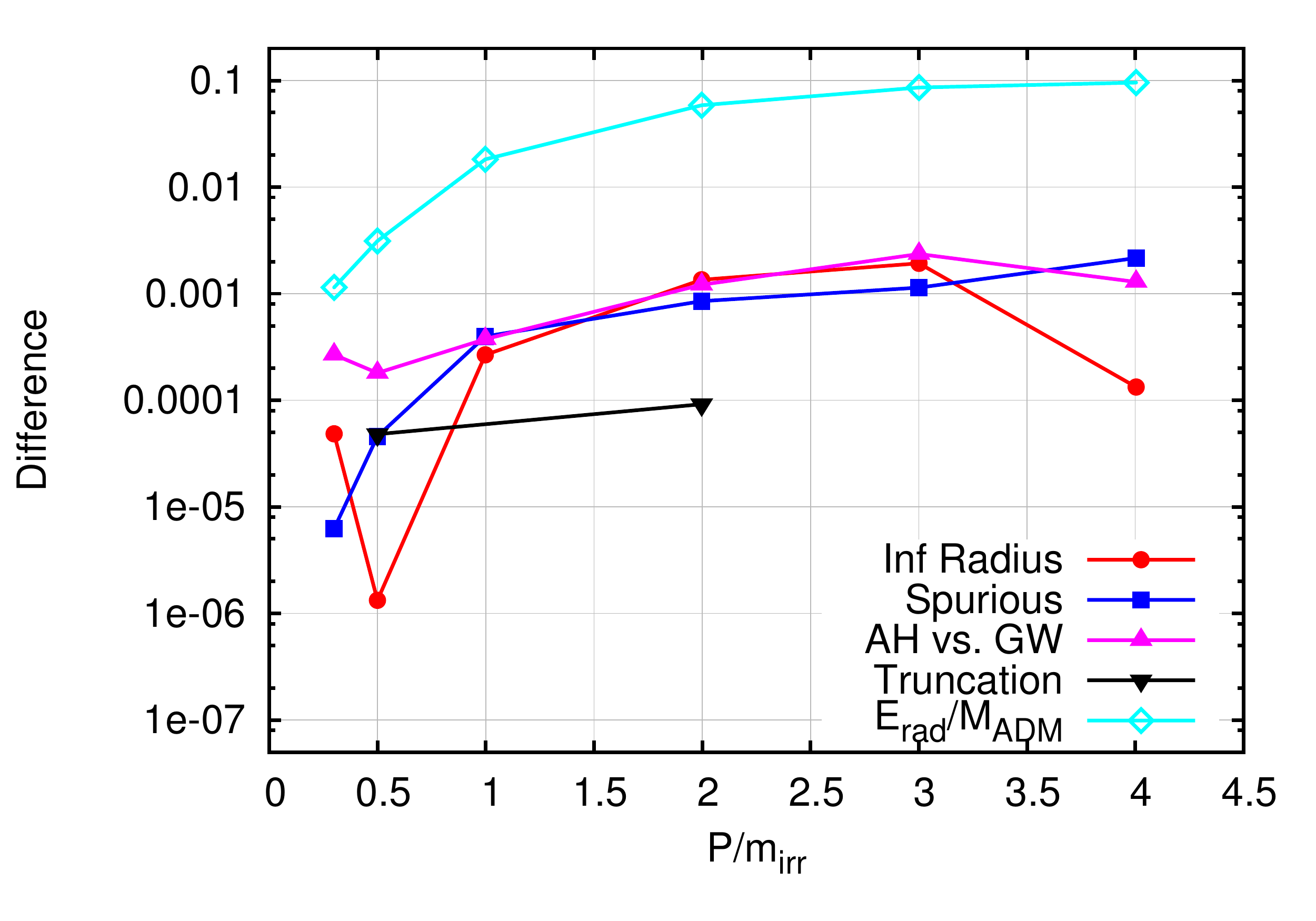}
\includegraphics[angle=270,width=\columnwidth]{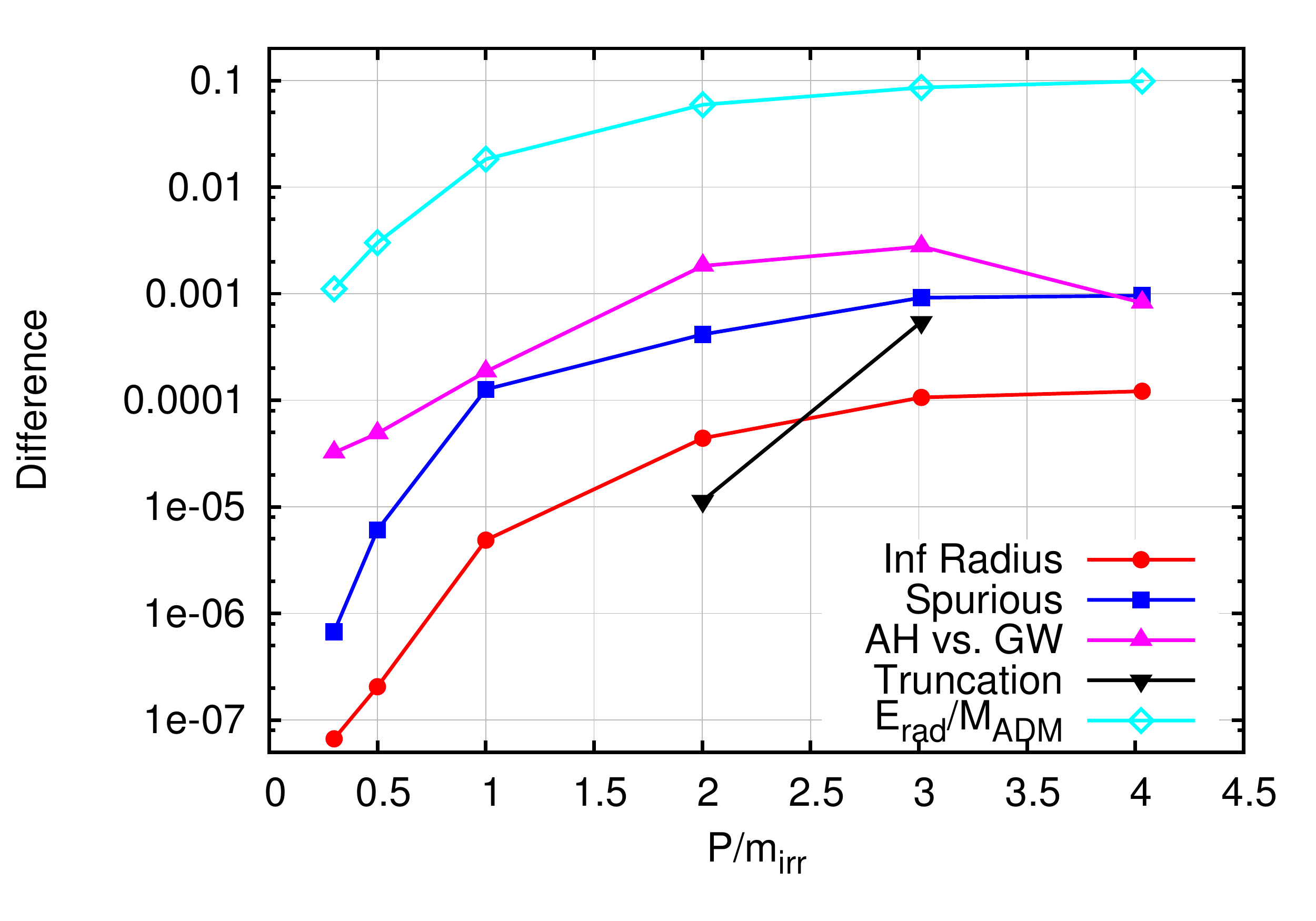}
\caption{
Estimated errors for each component of the computation of the radiated
energy. ``Inf Radius'' is the extrapolation from the finite extraction radius
$r_{\text{obs}}$ to infinity. ``Spurious'' is the effect of the initial radiation content
of the data. ``Truncation'' is an estimate of the finite difference
resolution used in the simulation, and ``AH vs. GW''
is a consistency measure of the radiated energy as computed by the
gravitational waveforms or the remnant mass of the final black hole.  Shown in cyan is
the total energy radiated for that simulation.
Top panel is the BSSNOK evolutions of the {\it standard data}, and
bottom panel is for the CCZ4 evolutions of the {\it approximate data}.
}
\label{fig:CCZ4_LB_hiro_errors}
\end{figure}

For the {\it standard data} simulations (which were all evolved using
BSSNOK),
the extraction radii extended to $r_{\text{obs}}=130M$.  
 The error in the extrapolation is 
estimated by the difference between the two fits and is labeled
``Inf Radius'' error in Fig.~\ref{fig:CCZ4_LB_hiro_errors}.

For the simulations of the {\it approximate data} (which were all
evolved using CCZ4), we needed to use larger initial separations than
for the {\it standard data} in order to reduce the constraint violations on
the initial slice. We therefore
extracted the
radiation at correspondingly large distances. For example, for the
largest separation run $d=400M$, the largest extraction radius as
$275M$ (note the black holes were initially located at $x=\pm200M$).

 In all of the CCZ4 cases, the extrapolation formula 
of Ref.~\cite{Nakano:2015pta} to $\mathcal{O}(1/r^2_{\text{obs}})$
gives a very robust set of
values for the radiated energy. To provide a generous bound, we used those two radii as 
estimates of the infinite radius energy radiated.

The other source of error we seek to keep under control is the
initial, unphysical radiation content.
We checked this spurious radiation for all of the Lorentz boosted
runs.
To do this, we compare the radiated energy of the full waveform
with that obtained by removing the initial transient.
The effect of the initial transient is to change the total
radiated energy by $\sim 1.6\%$ (relative to the total radiated
energy). The effect of this spurious radiation
on the accuracy of the total radiated energy is shown in 
Fig.~\ref{fig:CCZ4_LB_hiro_errors} under the label ``Spurious''.

It is worth noting here that the waveforms are extracted by
a multipole decomposition at the observer location. In 
practice a few of the lower modes are necessary for an accurate
account of the total radiation.
For instance, the $\ell$-mode contributions to the CCZ4
simulations for a $P/m_{\text{irr}}=3$ run (with initial separation $d=100M$) at
$r_{\text{obs}}=275M$ gives that $\ell=2$ contains $90\%$, 
$\ell=4$ contains $8.3\%$, and $\ell=6$ contains $1.68\%$ of the total 
energy radiated. Thus our results will include modes up to $\ell=6$.

\begin{figure}
  \includegraphics[angle=270,width=0.9\columnwidth]{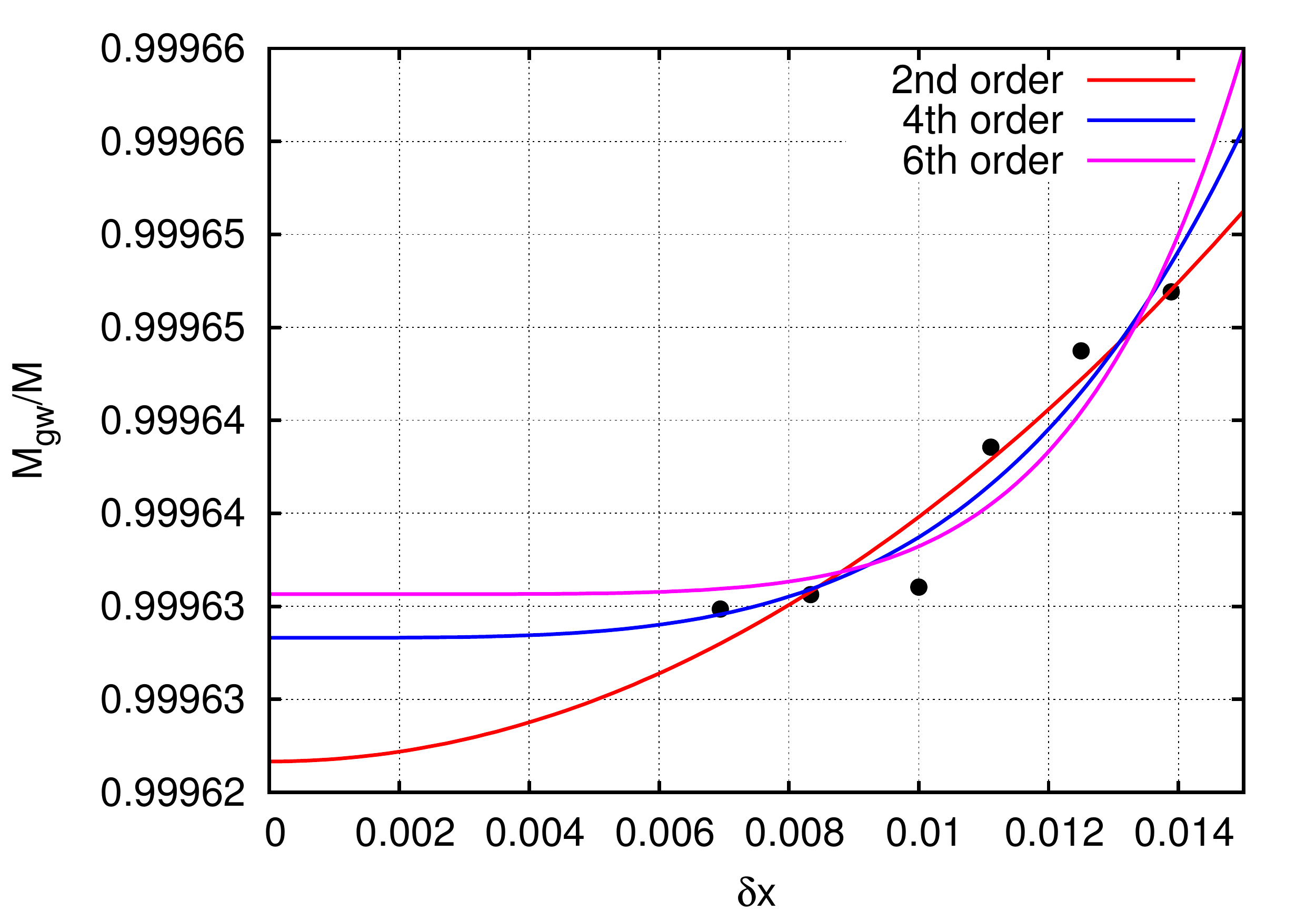}
  \caption{The convergence for the $P/m_{\text{irr}}=0.50$ final mass 
calculated from the gravitational modes $(\ell_{max}=6)$.  
sixth-, fourth-, and second-order fits to the data are shown. The fourth-order
fit is closest to the data.
For this convergence study,  we used six runs $(n72,n80,n90,n100,n120,n144)$.
}
  \label{fig:mgw_conv_p050_dO}
\end{figure}

To test the accuracy and consistency of our simulations, we 
performed a convergence study of the radiated energy, the main
physical quantity studied here, for six runs (all with 
initial $P/m_{\text{irr}}=0.5$). We increases the resolution in
stepsizes of $1.2$ between each run. 
The results of evaluation of
the final mass of the black hole from the measurement of the gravitational
radiation losses is shown in Fig.~\ref{fig:mgw_conv_p050_dO}.
While the differences with resolution are small, they are compatible
with the expected 4th order convergence of the evolution system.
In this figure, we fit the data to the form $y=a_0 + a_1 x^p$, where
$a_0$ and $a_1$ are fitting constants and $p$ is taken to be 2, 4, and
6.

\begin{figure}
  \includegraphics[width=0.9\columnwidth]{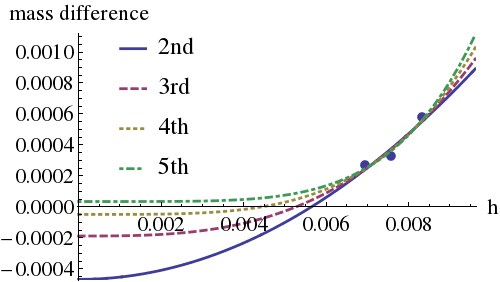}
  \caption{
The difference between the final horizon mass as calculated
    using the IH formalism and as inferred from the radiated energy for a
    $P/m_{\text{irr}}=2$ simulation as a function of resolution.
    The three highest resolution runs are shown. The data points
    themselves appear to be convergent. We estimate the infinite
    resolution limit by assuming second, third, fourth-order, and
    fifth-order
    convergence. The agreement between
    the horizon derived mass and radiation inferred mass at infinite
  resolution is $5\times 10^{-5}M$ for the expected 4th order convergence.}
  \label{fig:mass_conv}
\end{figure}

Also for runs with initial $P/m_{\text{irr}}=2$, 
the agreement between the
radiated energies, as measured from the waveforms (extrapolated to
observer location to infinity via~\cite{Nakano:2015pta}) and those
inferred from the initial ADM mass minus the remnant horizon mass,
provides a consistency 
measure for the numerical simulation (here we increased the resolution
by factors of 1.1). Assuming the differences scales
like $a\,h^b$, the $b-$power of convergence for the three highest resolution
$h$ runs is found to be $4.17697\pm1.139$. In addition,
we fit the same data to the form $a\,h^b + c$, where $b$ was fixed to
1, 2, 3, 4, and 5, and for each choice of $b$, we fit to $a$ and
$c$. The results are summarized
in Fig.~\ref{fig:mass_conv}, which shows different orders of extrapolation
to infinite resolution. The best results are  near the expected
4th order convergence. Here, we find consistency in the
final mass to within $5\times 10^{-5}M$.

Another interesting aspect to explore is how appropriate the standard
moving puncture gauges~\eqref{eq:gauge} are for evolving
highly-boosted black holes.
We found that at relatively short initial distances
the BSSNOK formalism generates a gauge wave focused by the
two black holes that then induces a large change in the lapse. This can drive
the lapse beyond $\alpha=1$ and trigger a Courant violation.
This problem
was resolved by starting the black holes at larger initial separations,
allowing the large gauge waves to sufficiently dissipate before the collision.
We also
found it beneficial
to use an initial lapse of the form
$\alpha_0=1/(2\psi_{\text{BL}}-1)$ and the approximate shock
avoiding gauge profile $f(\alpha)=8/(3 \alpha(3 - \alpha))$ (which we
used for all CCZ4 simulations).

\begin{figure}
\includegraphics[angle=270,width=.9\columnwidth]{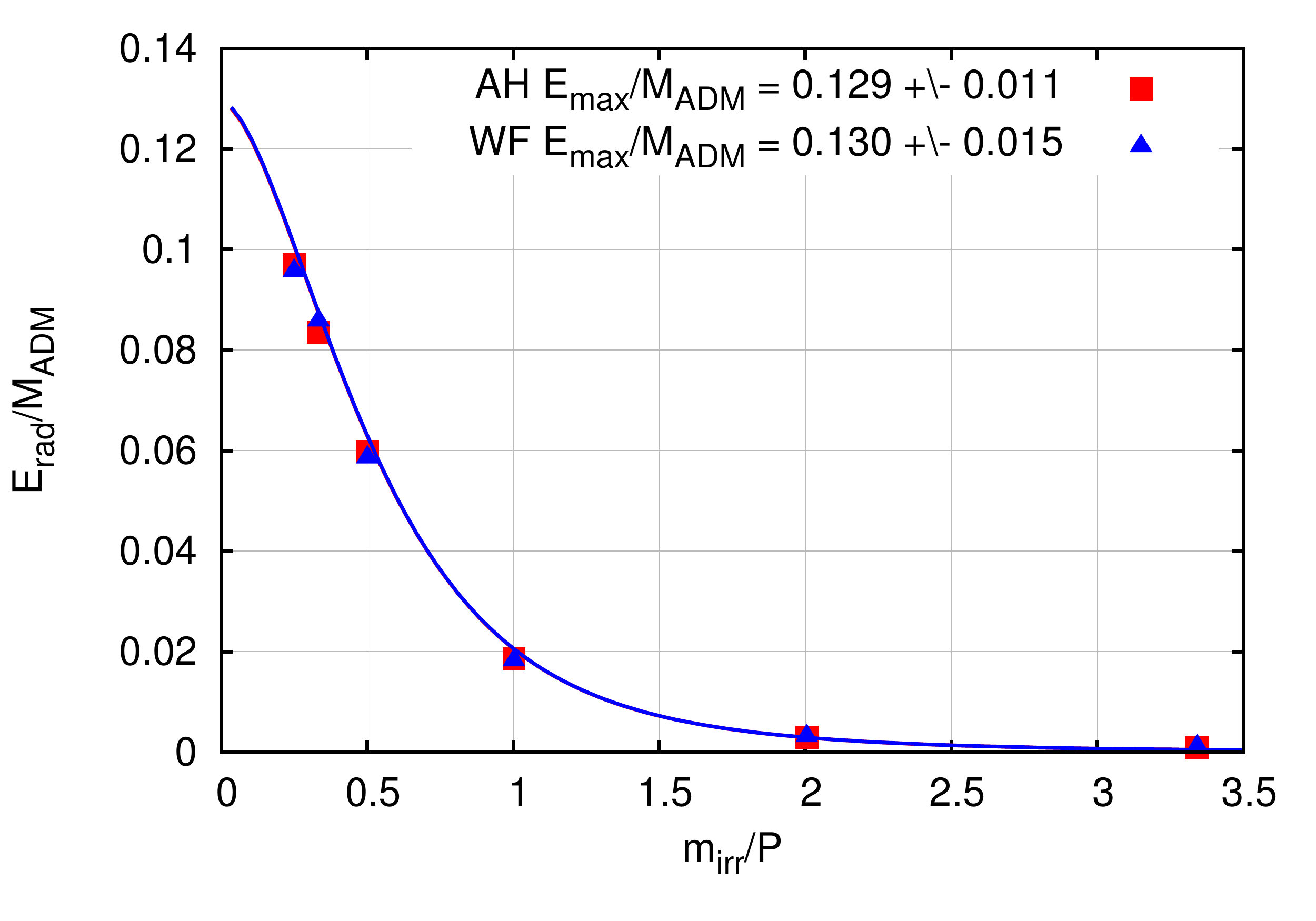}
\includegraphics[angle=270,width=0.9\columnwidth]{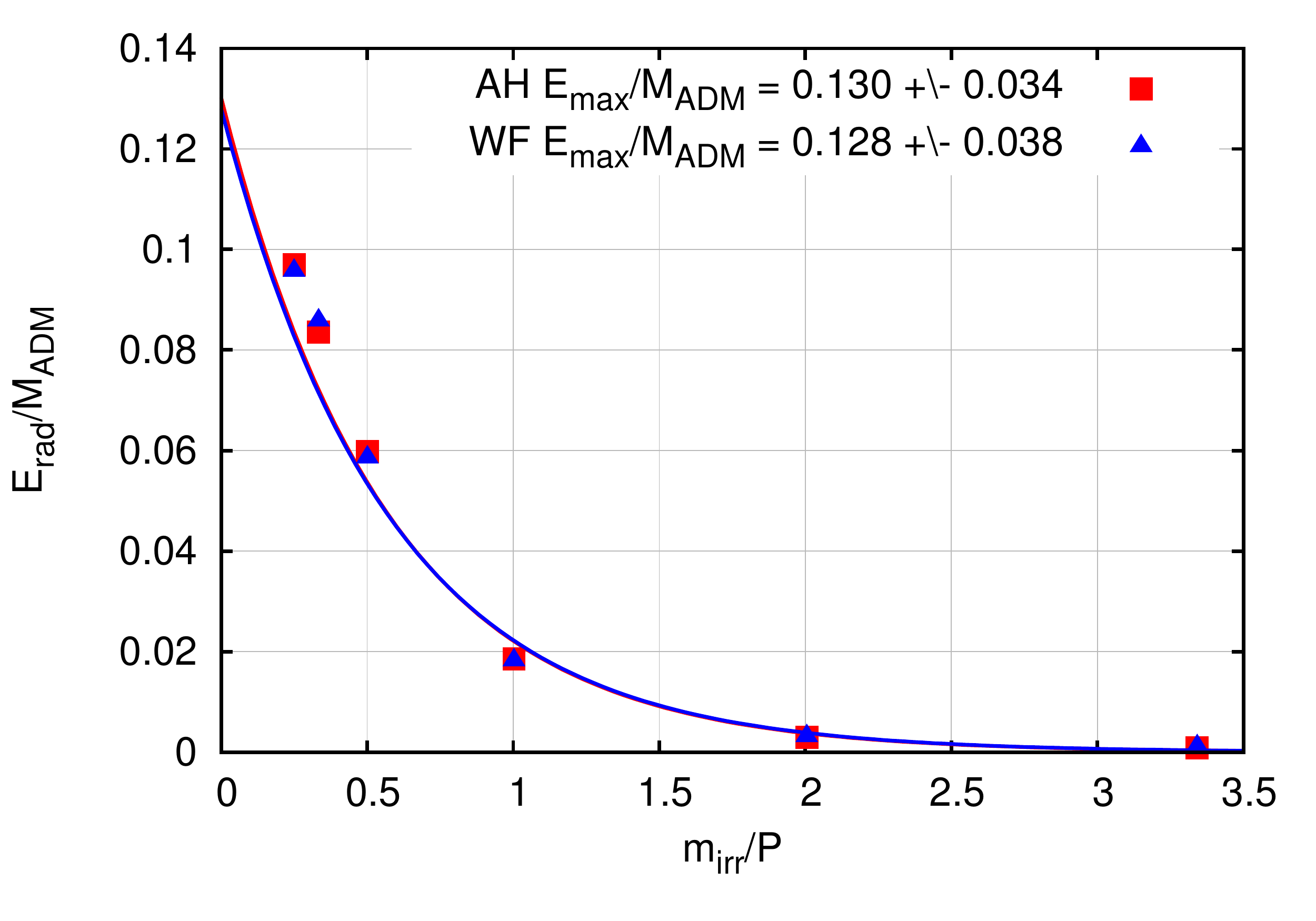}
\caption{Fits to the energy radiated at infinity for the 
  BSSNOK evolutions of the {\it standard data} in Table~\ref{tab:EradTab}.
Upper plot: Fit using the 1-parameter ZFL like fit.
Lower plot: An alternative two-parameter ($A$ and $B$) fit of the form
$(y=A\exp[-B\,x])$.  Both fits use data assuming a weighting error of the points of
$1\%$ and include fits to both the  energy radiated as measured by extraction
of radiation (WF) or by the remnant mass (AH).
}
\label{fig:BSSNEvsMoverP_fits}
\end{figure}

\begin{figure}
\includegraphics[angle=270,width=0.9\columnwidth]{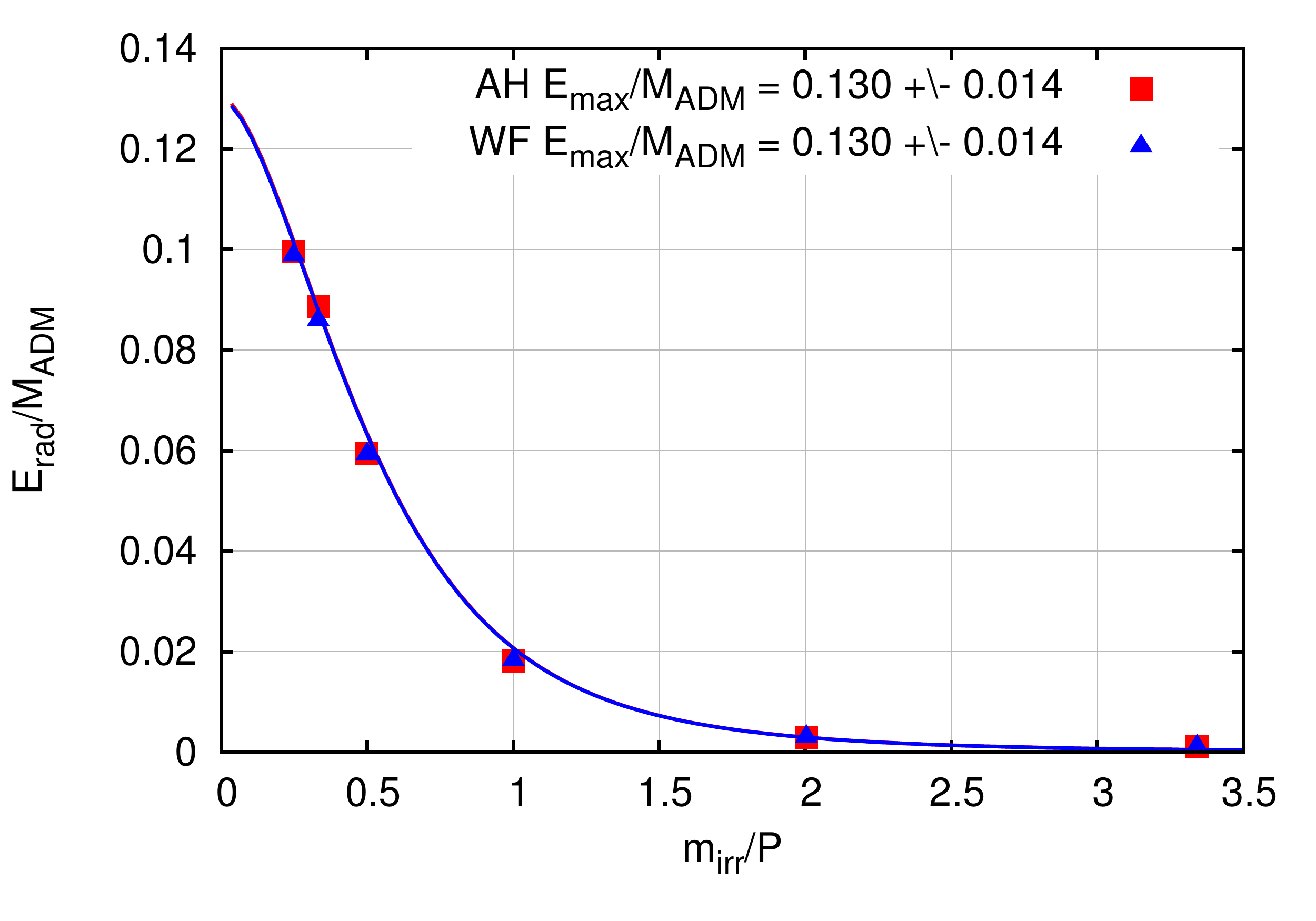}
\includegraphics[angle=270,width=0.9\columnwidth]{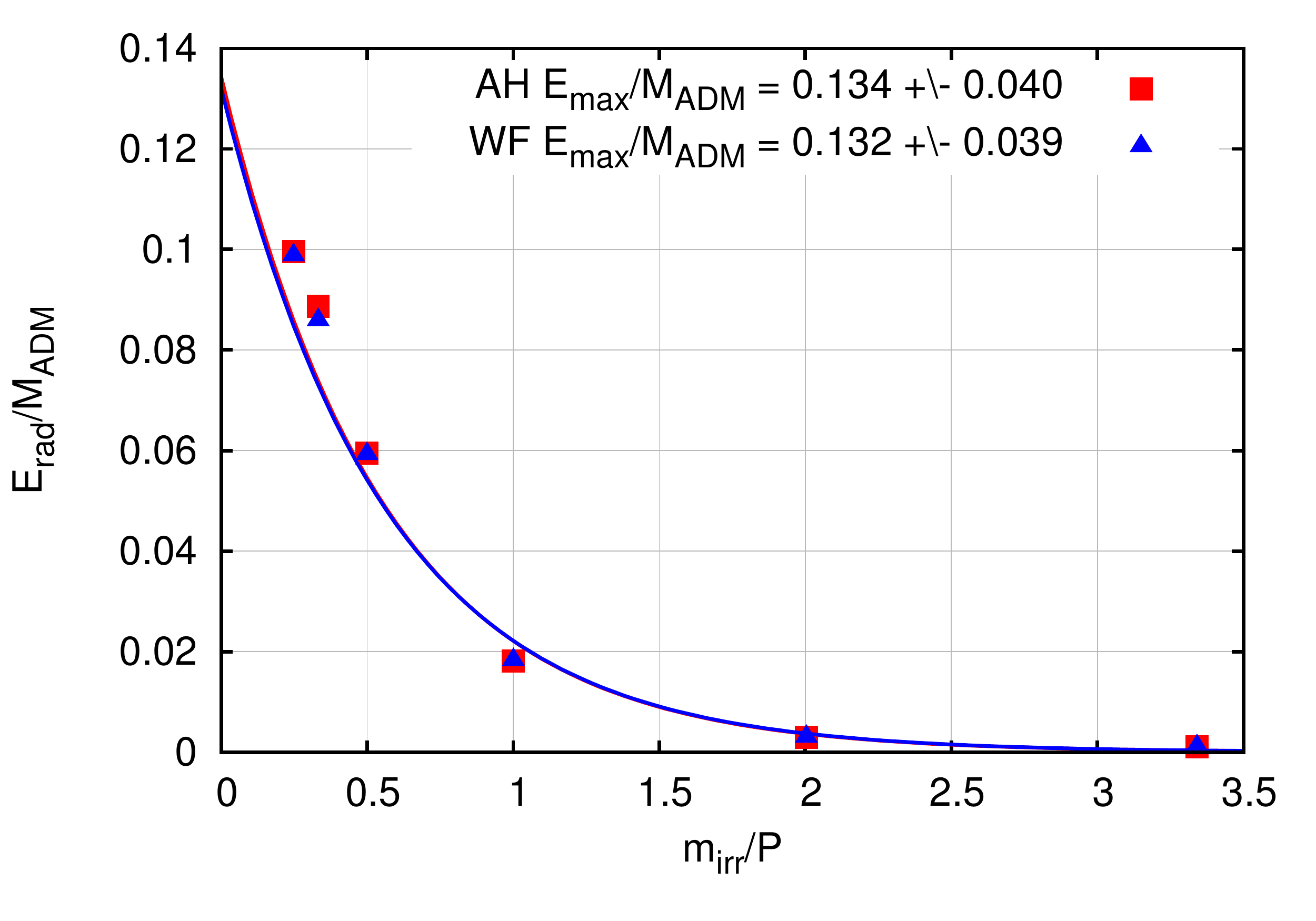}
\caption{Fits to the energy radiated at infinity for the CCZ4
  evolutions of the {\it approximate data} in Table~\ref{tab:EradTab2}.
Upper plot: Fit using the 1-parameter ZFL like fit.  
Lower plot: An alternative two-parameter ($A$ and $B$) fit of the form
$(y=A\exp[-B\,x])$.  Both fits use data assuming a weighting error of the points of 
$1\%$ and include fits to both the  energy radiated as measured by extraction 
of radiation (WF) or by the remnant mass (AH). 
}
\label{fig:EvsMoverP_fits}
\end{figure}

When fitting the radiated energy as a function of the initial
momentum we assume a relative error of $1\%$
in each computed energy to determine its weight in the fit.
We fit these energy values as a function of the variable
$m_{\text{irr}}/P$, where $m_{\text{irr}}$ stands for
the irreducible mass of each initially boosted black hole with momentum
$\pm P$. The upper panels in 
Fig.~\ref{fig:BSSNEvsMoverP_fits} and~\ref{fig:EvsMoverP_fits} display the results of the fitting,
assuming the dependence of the energy radiated is given by the ZFL
behavior~\cite{Smarr:1977fy,Sperhake:2008ga}
\begin{equation}
  \label{eq:ZFL}
  \frac{E}{M}=E_{\infty}\left(\frac{1+2\gamma^2}{2 \gamma^2}+\frac{(1-4\gamma^2)\log{(\gamma+\sqrt{\gamma^2-1})}}{2 \gamma^3\sqrt{\gamma^2-1}}\right) \; ,
\end{equation}
where $\gamma=\sqrt{1+(P/m_{\text{irr}})^2}$ and 
the only fitting parameter is $E_{\infty}$.
The relative deviations
are mostly below $10\%$, and in particular are around $2\%$ for the most energetic simulated collision.

To assess the dependence with the chosen fitting function,
we have assumed a fit of the form $(y=A\exp[-B\,x])$ with two
fitting parameters ($A$ and $B$), $y$ and $x$ being the independent and
dependent variables, i.e. $E_{\text{rad}}/M_{\text{ADM}}$ and $m_{\text{irr}}/P$, respectively.
The results of this fit are displayed in the lower panels of
Fig.~\ref{fig:BSSNEvsMoverP_fits} and~\ref{fig:EvsMoverP_fits}. In spite of introducing two
fitting parameters, we observe that the residuals are larger than the fit
using the ZFL form~\eqref{eq:ZFL}, thus
rendering further support to this behavior.
We have also experimented with fittings of the form
$(y=A\exp[-B\,x^C])$, introducing a third parameter $C$ in the fitting
function, and also assuming $C=2$, but none of these options displayed
better behavior than the ZFL choice.

In either case of the fits shown in Fig.~\ref{fig:EvsMoverP_fits}, the
estimated maximum radiated energy is around $13\%$, which provides
a robust estimate, all errors considered, of the form 
$E_{\text{max}}/M_{\text{ADM}}=0.13\pm0.01$.

\section{Conclusions and Discussion}\label{sec:discussion}

Using improved full numerical techniques, we have been able to
provide a more accurate determination of the maximum gravitational
radiation produced in the head-on collision of nonspinning black holes.
These techniques utilize initial data for highly boosted black holes~\cite{Ruchlin:2014zva}
with much less radiation content than the Bowen-York counterparts, and
reach near the ultrarelativistic regime with speeds much closer to $c$.
We have successfully extrapolated the extracted waveforms
to infinite observer locations with the techniques of 
Ref.\@~\cite{Nakano:2015pta}, and added up to $\ell=6$ 
modes in the computation of the radiated energy.
The evolutions of the initial data have been carried out using the
\MP approach using both the BSSNOK and CCZ4 systems.

We find a maximum radiated energy
of $13\pm1\%$ of the total mass of the system, with most of the
errors coming from the functional fitting and subsequent extrapolation to infinite 
boost. This result is in close agreement
with the analytic estimates of $13.4\%$ of Ref.\@~\cite{Siino:2009vw}
using thermodynamic arguments and the previous numerical estimate of
$14\pm3\%$ in Ref.\@~\cite{Sperhake:2009jz}. However, they seem to be in conflict with the analytic estimates
of $16.4\%$ from second order perturbations~\cite{D'Eath:1992qu}
and $17\%$ from the multipolar analysis of the
ZFL~\cite{Berti:2010ce}.\\

\acknowledgments 
The authors thank M.\ Campanelli for helpful discussions.
The authors gratefully acknowledge the NSF for financial support from Grants
PHY-1305730, PHY-1212426, PHY-1229173,
AST-1028087, PHY-0969855, OCI-0832606, and
DRL-1136221. Computational resources were provided by XSEDE allocation
TG-PHY060027N, and by NewHorizons and BlueSky Clusters 
at Rochester Institute of Technology, which were supported
by NSF grant No. PHY-0722703, DMS-0820923, AST-1028087, and PHY-1229173.
This work was supported in part by the 2013--2014 Astrophysical Sciences and Technology Graduate Student Fellowship,
funded in part by the New York Space Grant Consortium, administered by Cornell University.

\bibliographystyle{apsrev4-1}
\bibliography{../../../../Bibtex/references}

\end{document}